\documentclass[aps,showkeys,showpacs,tightenlines,nofootinbib,prd]{revtex4}
\usepackage{amsmath,amscd,amssymb}
\usepackage{color,epsfig}
\usepackage{xfrac}
\usepackage{latexsym}
\usepackage{mathrsfs}
\usepackage{graphicx}
\usepackage{subfigure}
\usepackage{ulem}
\usepackage{bbm}
\newcommand{\imu}{{\rm i}}

\DeclareMathOperator\sech{sech}

\begin{document}

\title{Vacuum Polarization Energy of the Kinks in the Sinh-Deformed Models.}

\author{I. Taky, B. Barnes, J. Ackora-Prah}
\affiliation{Mathematics Department, Kwame Nkrumah University of Science and Technology, Kumasi, Ghana\\
Email: ishmael.takyi@knust.edu.gh and ishmael@aims.ac.za}

\begin{abstract}
We compute the one-loop quantum corrections to the kink energies of the sinh-deformed $\phi^{4}$ and $\varphi^{6}$ models in one space and one time dimensions. These models are constructed from the well-known polynomial $\phi^{4}$ and $\varphi^{6}$ models by a deformation procedure. We also compute the vacuum polarization energy to the non-polynomial function $U(\phi)=\frac{1}{4}(1-\sinh^{2}\phi)^{2}$. This potential approaches the $\phi^{4}$ model in the limit of small values of the scalar function. These energies are extracted from scattering data for fluctuations about the kink solutions. We show that for certain topological sectors with non-equivalent vacua the kink solutions of the sinh-deformed models are destabilized.
\end{abstract}


\maketitle

\section{Introduction}

Kinks are classical solutions to the non-linear field equations in one space and one time dimensions~\cite{Rajaraman:1982is}. An important property of the kink model is that in the classical picture it resembles an extended particle. Thus, it possesses localized energy densities. The kink solutions from these non-linear models behave particle like when subjected to extended forces and have been applied in many branches of physics: in cosmology~\cite{Vachaspati:2006zz, Vilenkin:2000jqa} the kink solutions describe the cosmic domain walls~\cite{Anninos:1991un,Romanczukiewicz:2017hdu}, in condensed matter physics~\cite{bishop1980solitons} they are used to study Bose-Einstein condensates~\cite{kevrekidis2008condense} and ferromagnets~\cite{Ivanov:1992aa}, as well as particle physics~\cite{Weigel:2008zz}.

These kinks are referred to as topological kinks because they are characterized by a topological index which is related to their behavior at spatial infinity. A typical example is the Skyrmion a model for the baryon ~\cite{Skyrme:1988xj,Adkins:1983ya,Weigel:2008zz}. In this regime, the topological index becomes a conserved quantum number called the baryon number. In one space and one time dimensions, kink solutions interpolate neighboring vacua at negative and positive spatial infinity, where the masses of the quantum fluctuations about the degenerate vacua differ. In this project, we investigate field theories, whose classical solutions connect vacua with different masses of quantum fluctuations. 

In the Skyrme model, the integral of the energy density is inversely proportional to its coupling constant and is identified as the mass of the particle. This mass overestimates the actual mass of the particle on the grounds that, quantum corrections are ignored ~\cite{Moussallam:1991rj,Holzwarth:1995bv,Meier:1996ng}. This is not a problem when investigating the properties of a single particle. In computations of the binding energies of compound objects such as hypernuclear atoms~\cite{Meoto:2019hdi}, the quantum corrections may become important when comparing configurations with different particle numbers. The vacuum polarization energies (VPE) are the leading quantum correction to the kink energies and are the renormalized sum of the shifts of zero points energies of the quantum fluctuations due to their interactions with the (classical) background potential. We compute the VPE of the kink of sinh-deformed potentials using the spectral methods~\cite{graham:2009spec}.

The VPE has been investigated by several researchers, for example, the kinks in the $\phi^{4}$~\cite{Dashen:1974cj}, and sine-Gordon models~\cite{Dashen:1975hd,Faddeev:1977rm}, and cosmic strings in the standard model~\cite{graham:2009spec,Achucarro:1999it}. Recent studies in the $\varphi^{6}$ model~\cite{Weigel:2017iup,Weigel:2016zbs} revealed that the VPE destabilized the kink as the kink produces different curvatures for the quantum fluctuations at both positive and negative spatial infinity. These instabilities of the kink have also been observed in the $\phi^{8}$ model~\cite{Takyi:2020tvl}. 

The non-polynomial models are another aspect where kink has played a major role. In Refs. \cite{Bazeia:2017rxo,Bazeia:2019xoe}, the authors observed a pattern of kink-antikink scattering of the sinh-deformed models which was consistent with the observation in polynomial models of the same order~\cite{Takyi:2016tnc,Weigel:2013kwa,Anninos:1991un,Campbell:1983xu,Dorey:2011yw}. The study of this pattern amounts to numerically solving the equations of motion for time and space dependent fields with specific initial conditions: in the distant past, the kink and antikink are well separated and do not interact. When boosted with a prescribed velocity, the kink and antikink approach each other and interact at a later stage. This velocity is referred to as relative velocity between the kink and antikink. The patterns are observed for certain values of the relative velocity below a specific critical velocity. The remarkable physical phenomenon observed in their scattering analysis makes it interesting to further investigate other physical properties of it.  

In this paper, we consider the non-polynomial hyperbolic potential of the $\phi^{4}$ and $\varphi^{6}$ types. We compute the VPE of the kink potentials in one space and one time dimensions for these models. These models have kink solutions similar to the polynomial $\phi^{4}$ and $\varphi^{6}$ models with spontaneous symmetry breaking. For this reason, we will compare the VPE results to the polynomial $\phi^{4}$ and $\varphi^{6}$ models to highlight their differences. The authors in Ref.~\cite{Bazeia:2019xoe} numerically studied the kink structures of the potential $U(\phi)=\frac{1}{4}(1-\sinh^{2}\phi)^{2}$. This potential mimic the $\phi^{4}$ potential in the limit as the scalar function becomes small. We numerically compute the quantum corrections for this model and compare them to its polynomial counterpart.

We organized our work as follows: In the next section, we briefly review the general properties of static kinks with finite energy in one space and one time dimensions. We will review the method of computing the VPE in Section \ref{vpe}. In Section \ref{model} we introduce the models we consider and present the numerical results in Section \ref{results}. We conclude in Section \ref{conc}.

\section{Kink Concept}
\label{kinkconcept}
We consider a single scalar field $\phi(x,t)$ in one space $(x)$ and one time $(t)$ dimensions, whose dynamics is defined by the Lagrangian density 
\begin{equation}
	\mathcal{L} = \frac{1}{2} \left(\phi_{t}\right)^{2} - \frac{1}{2}\left(\phi_{x}\right)^{2} - U (\phi),
	\label{eq:Lang}
\end{equation}
where the subscripts, $x$, $t$, denote differentiation with respect to $x$ and $t$, respectively.
Here, $U(\phi)$ is the quantum field potential with two or more degenerate minima. The corresponding field equation for the Lagrangian is
\begin{equation}
	\phi_{tt}-\phi_{xx}= - \frac{{\rm d}U}{{\rm d} \phi}.
	\label{eq:ELang}
\end{equation}
For static configuration, $\phi_{t}=0$ and Eq.(\ref{eq:ELang}) reduces to 
\begin{equation}
	\phi_{xx}=\frac{{\rm d}U}{{\rm d} \phi}.
	\label{eq:staticequa}
\end{equation}
For finite energy, Eq.(\ref{eq:staticequa}) transforms into a first-order differential equation 
\begin{equation}
	\frac{{\rm d}\phi}{{\rm d}x} = \pm \sqrt{2U(\phi)}.
\end{equation}
In this case, the kink mass (classical energy) is given by
\begin{equation}
	E[\phi] = \int \sqrt{2U(\phi)} {\rm d} \phi,
\end{equation}
where the integration boundaries are two neighboring potential minima.

To analyse the linear stability of the kink, we call the kink solution to the field equation $\phi_{K}$ and parametrize the field 
\begin{equation*}
	\phi(x,t)=\phi_{K}(x) + \eta(x,t).
\end{equation*}
By considering linear terms in $\eta$ we get a Schr\"odinger-like equation in one dimension
\begin{equation}
	\left[ -\partial^{2}_{x} + u(x) \right] \eta(x) = \omega^{2} \eta(x),
	\label{eq:waveequa}
\end{equation}
where $\eta(x)$ and $\omega$ are eigenfunctions and eigenvalues of the Schr\"odinger operator $\displaystyle H:=  -\partial^{2}_{x} + u(x) $ and 
\begin{equation}
	u(x) = \left. \frac{{\rm d}^{2} U}{{\rm d} \phi^{2}} \right|_{\phi_{K}(x)}
\end{equation}
is the scattering potential which is generated by the background kink. At positive and negative spatial infinity for non-equivalent vacua, $u(x)$ approaches a constant (so-called meson masses), i.e $\lim_{x \rightarrow +\infty} u(x) = m_{R}^{2}$ and $\lim_{x \rightarrow -\infty} u(x) = m_{L}^{2}$. For non-equivalent vacua we take $m_{L} \leq m_{R}$. The analyses of the scattering potential sheds light on the scattering structure of the kink.

The sinh-deformed models are obtained from the known potentials $U(\phi)$ to another potential $V(\phi)$ by a deforming function $g(\phi)=\sinh \phi$~\cite{Bazeia:2006pj,Bazeia:2002xg,Almeida:2004ai,Bazeia:2005hu,Bazeia:2017nlo}, 
\begin{equation}
	V(\phi) = \frac{U\left( g(\phi)\right)}{\left(g^{\prime}(\phi)\right)^{2}}.
	\label{eq:deform_proc}
\end{equation}
The static solutions for the new potential $V(\phi)$ is given by 
\begin{equation}
	\phi_{K}^{\rm (new)}(x)=g^{-1}\left(\phi_{K}^{\rm (old)}(x)\right),
	\label{eq:defmakink}
\end{equation}
where $\phi_{K}^{\rm (old)}(x)$ is the static solution for the known potential $U(\phi)$.

\section{Vacuum Polarization Energy}
\label{vpe}
The VPE, $E_{\rm vac}$ is the leading quantum correction to the classical kink energy. It is the renormalized sum of the shifts of the zero-point energies of the quantum fluctuations due to their interaction with the background configuration generated by the kink. Formally the VPE reads
\begin{equation}
	E_{\rm vac} = \frac{1}{2} \sum_{j}^{\rm b.s} \omega_{j} + \frac{1}{2} \int_{0}^{\infty} {\rm d}k\, \omega_{k} \Delta \rho(k) + E_{\rm ct}
	\label{eq:vpe1}
\end{equation}
where the first term on the right hand side of Eq. (\ref{eq:vpe1}) is the contribution from the discrete bound states $({\rm b.s})$ $\omega = \omega_{j}$ with $\vert \omega_{j} \vert \leq m_{L}$ and the second term, is the continuum contribution weighted by the change in the density of states, $\Delta \rho(k)$. The third term on the right hand side of Eq. (\ref{eq:vpe1}), $E_{\rm ct}$ is contribution from the counterterms that yields a finite results at one-loop level. Here $\omega_{k}=\sqrt{k^{2} + m_{L}^{2}}$ are the energies of the scattering states where $k$ is the momentum and $m_{L}$ is the mass of the mesons. 

The modified density states $\Delta \rho(k)$, is measured by the derivative of the scattering eigen-phase shifts, extracted from the scattering matrix $S(k)$
\begin{equation}
	\Delta \rho(k)=\frac{1}{\pi}\frac{{\rm d}}{{\rm d}k} \delta(k), \quad \delta(k)=-\frac{\imu}{2} \ln \left[\det S(k)\right].
	\label{eq:phaseshift}
\end{equation}
We compute $S(k)$ by introducing a pseudo-potential 
\begin{equation}
	u_{p}(x)=u(x)-m_{L}^{2} + \left(m_{L}^{2}-m_{R}^{2}\right)\Theta \left(x-x_{m}\right) 
	\label{eq:pseudopot}
\end{equation}
which vanishes at positive and negative spatial infinity. Here, $\Theta(x)$ is the step function and $x_{m}$ is an arbitrary matching point. Then the wave-equation becomes
\begin{equation}
	\left[-\partial^{2}_{x}+u_{p}(x)\right] \eta(x) =
	\begin{cases}
		k^{2} \eta(x), & \quad \text{for} \quad x \leq x_{m} \\
		q^{2} \eta(x), & \quad \text{for} \quad x \geq x_{m}
	\end{cases}
\end{equation}
where, $q=\sqrt{\omega^{2}-m_{R}^{2}}$. Above threshold $q$ is real so we take $k \geq \sqrt{m_{R}^{2}-m_{L}^{2}}$ and formulate a variable phase approach \cite{cal1987} by parametrizing
\begin{align}
	x \leq x_{m}: & \eta(x)=A(x)e^{\imu kx}, \quad A^{\prime \prime}(x) = -2 \imu k A^{\prime}(x) + u_{p}(x)A(x) \nonumber \\
	x \geq x_{m}: & \eta(x)=B(x)e^{\imu qx}, \quad B^{\prime \prime}(x) = -2 \imu q B^{\prime}(x) + u_{p}(x)B(x)
	\label{eq:parametequa}
\end{align} 
where a prime denotes a derivative with respect to $x$. The boundary condition $B(\infty)=A(-\infty)=1$ and $B^{\prime}(\infty)=A^{\prime}(-\infty)=0$ solve Eq.(\ref{eq:parametequa}) yielding the scattering matrix 
\begin{equation}
	S(k)=\begin{pmatrix}
		{\rm e}^{-iqx_m} & 0 \cr 
		0 & {\rm e}^{ikx_m}
	\end{pmatrix}
	\begin{pmatrix}
		B & -A^\ast \cr
		iqB+B^\prime & ikA^\ast-A^{\prime\ast}
	\end{pmatrix}^{-1}
	\begin{pmatrix}
		A & -B^\ast \cr
		ikA+A^\prime & iqB^\ast-B^{\prime\ast}
	\end{pmatrix}
	\begin{pmatrix}
		{\rm e}^{ikx_m} & 0 \cr 
		0 & {\rm e}^{-iqx_m}
	\end{pmatrix}\,,
	\label{eq:Smatrix}
\end{equation}
where $A=A(x_{m})$, etc. are the coefficient functions at the matching point. Below the threshold, i.e $k \leq \sqrt{m_{R}^{2}-m_{L}^{2}}$, $q=\imu \kappa$ becomes imaginary with $\kappa = \sqrt{m_{R}^{2}-m_{L}^{2}-k^{2}} \geq 0$. We parametrize the wave equation for $x \geq x_{m}$ as $\eta(x)=B(x) e^{-\imu \kappa x}$. This yields an ordinary differential equation 
\begin{equation*}
	B^{\prime \prime}(x)=\kappa B^{\prime}(x) + u_{p}(x)B(x).
\end{equation*}
We then extract the reflection coefficient via
\begin{equation}
	S(k) = -\frac{A\left(B^{\prime}/B- \kappa - \imu k\right)-A^{\prime}}
	{A^\ast\left(B^{\prime}/B-\kappa + \imu k\right)-A^{\prime\ast}} e^{2 \imu kx_m}\,.
	\label{eq:reflecoeff}
\end{equation}
The right hand side of Eq.(\ref{eq:reflecoeff}) have a negative sign which agrees with the statement of Levinson's theorem~\cite{Dong:1998vs}. This theorem relates the phase shifts of the scattered wave at infinity energy and zero energy, and it states that $\delta(0)$ is an odd multiple of $\frac{\pi}{2}$. It is given by 
\begin{equation}
	\delta(0)=\pi\left(n-\frac{1}{2}\right),
	\label{eq:levisonthe}
\end{equation}
where $n$ counts the total number of bound states.

Adopting the no-tadpole renormalization scheme \cite{Weigel:2016zbs}, the counterterm contribution in Eq.(\ref{eq:vpe1}) subtracts exactly the Born approximation $\delta^{(1)}$ from the phase shift \cite{graham:2009spec}. For non-equivalent vacua with different mesons masses, there is a direct contribution from the pseudo-potential as well as from the step function potential 
\begin{equation}
	\delta^{(1)}(k)=-\frac{1}{2k}\int_{-\infty}^\infty dx\, u_p(x)\Big|_{x_m}
	+\frac{x_m}{2k}\left(m_R^2-m_L^2\right)
	=-\frac{1}{2k}\int_{-\infty}^\infty dx\, u_p(x)\Big|_{0}\,,
	\label{eq:Born}
\end{equation}
where the subscript defines the position of the step in the pseudopotential $u_{p}(x)$. In the end, the Born approximation does not depend on $x_{m}$, even though $u_{p}(x)$ was initially defined in terms of $x_{m}$. We then obtain the total VPE as
\begin{equation}
	E_{{\rm vac}} = \frac{1}{2} \sum_{j}\left(\omega_{j}-m_L\right) -
	\frac{1}{2\pi} \int_{0}^{\infty} {\rm d}k\, \frac{k}{\sqrt{k^2+m_{L}^{2}}}
	\left(\delta(k)-\delta^{(1)}(k)\right)\,.
	\label{eq:evac}
\end{equation}
For reflection symmetric background potentials, the VPE is equivalently calculated by making use of analytic properties of scattering data and yields \cite{graham:2009spec}
\begin{equation}
	E_{{\rm vac}}^{S}=\int_{m_{L}}^{\infty} \frac{{\rm d}t}{2\pi} 
	\frac{t}{\sqrt{t^2-m_{L}^{2}}} \Bigg[\ln\Bigg\{ g(0,t) \Bigg( 
	g(0,t)-\frac{1}{t}g^{\prime}(0,t)\Bigg) \Bigg\}\Bigg]_{1\,.}
	\label{eq:Jost}
\end{equation}
The subscripts indicates that the Born approximation has been subtracted. The function $g(x,t)$ is the Jost solution factor on the imaginary axis that solves the differential equation 
\begin{equation}
	g^{\prime\prime}(x,t)=2tg^{\prime}(x,t)+V(x)g(x,t)
\end{equation}
with boundary conditions $g(\infty,t)=1$ and $g^{\prime}(\infty,t)=0$.

\section{ Models} 
\label{model}
Here we consider the models for our calculations. We make use of the natural units $\hbar = c =1$.
\subsection{The Sinh-Deformed Models}
\label{models}
\begin{figure}
	\centering
	\subfigure[Field potential of the sinh-deformed (dashed line) and polynomial (solid line) $\phi^{4}$ model. ]{\includegraphics[scale=0.3]{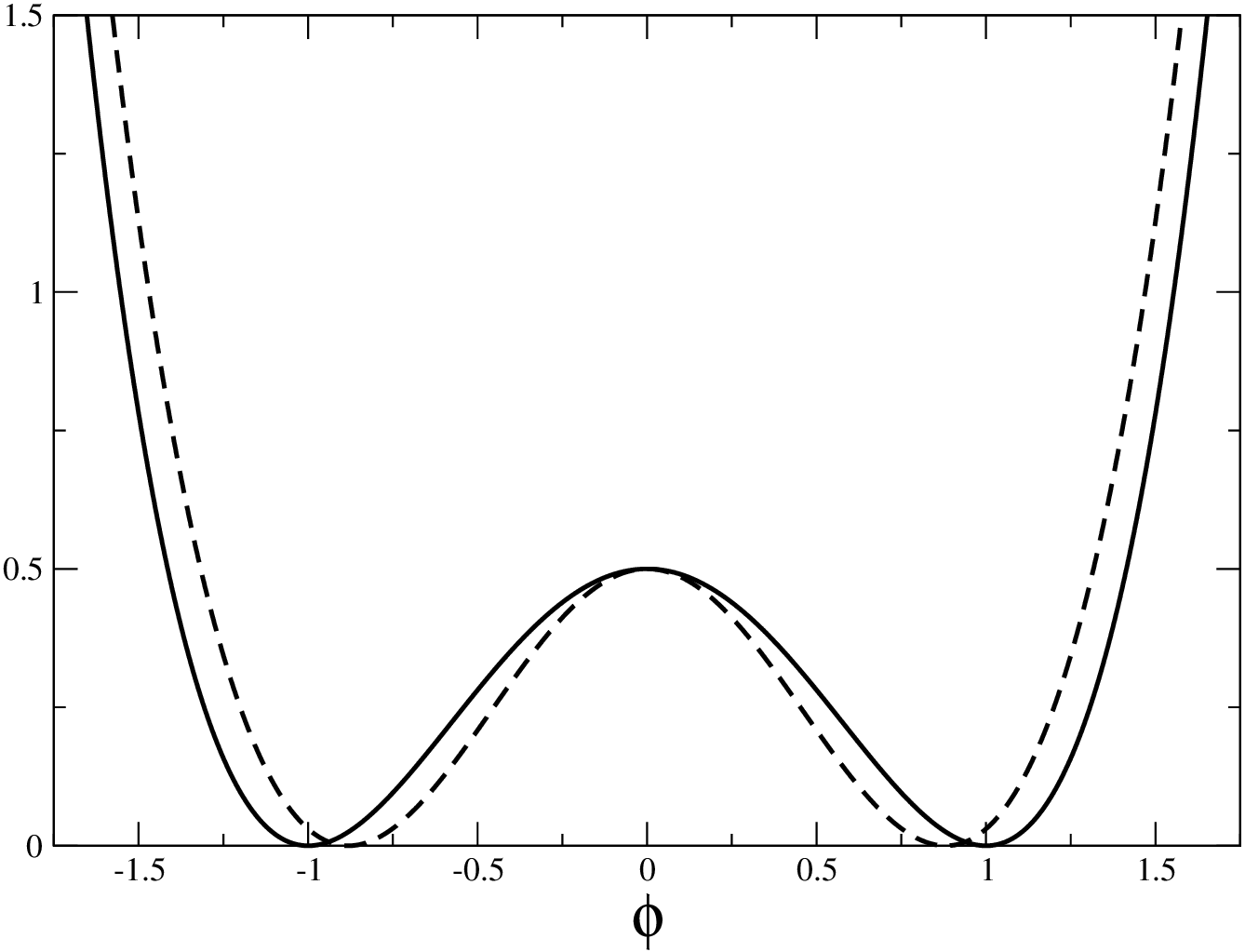}
		\label{fig:kinkinterphimodel_a}}
	\quad 
	\subfigure[Kink solutions of the sinh-deformed (dashed line) and polynomial (solid line) $\phi^{4}$ model.]{ 
		\includegraphics[scale=0.3]{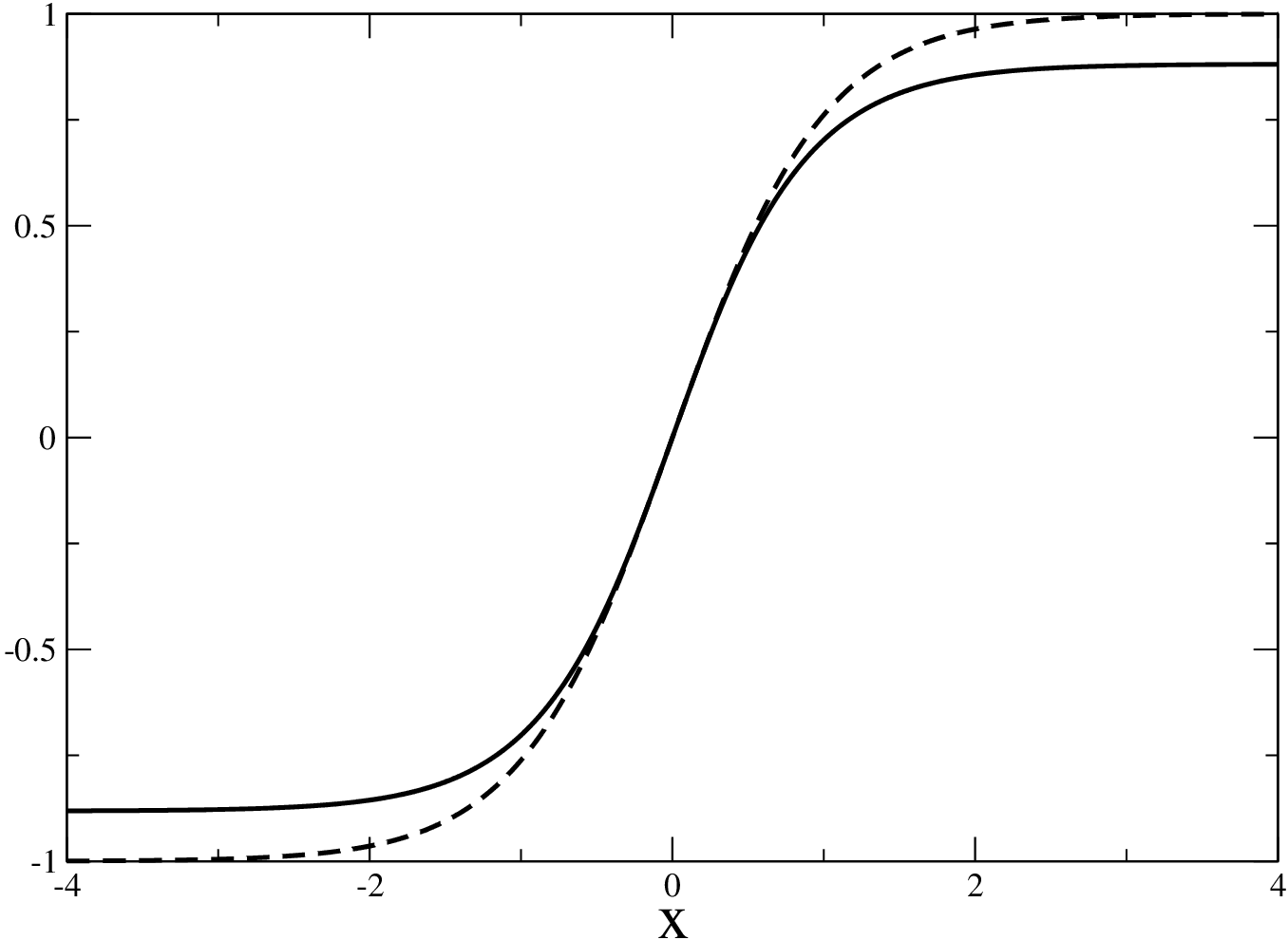}
		\label{fig:kinkinterphimodel_b} }
	\caption{\label{fig:kinkinterphimodel} The field potential and kink solutions of the $\phi^{4}$ model.}
\end{figure}

\begin{figure}
	\centering
	\subfigure[$a=0$ ]{\includegraphics[scale=0.22]{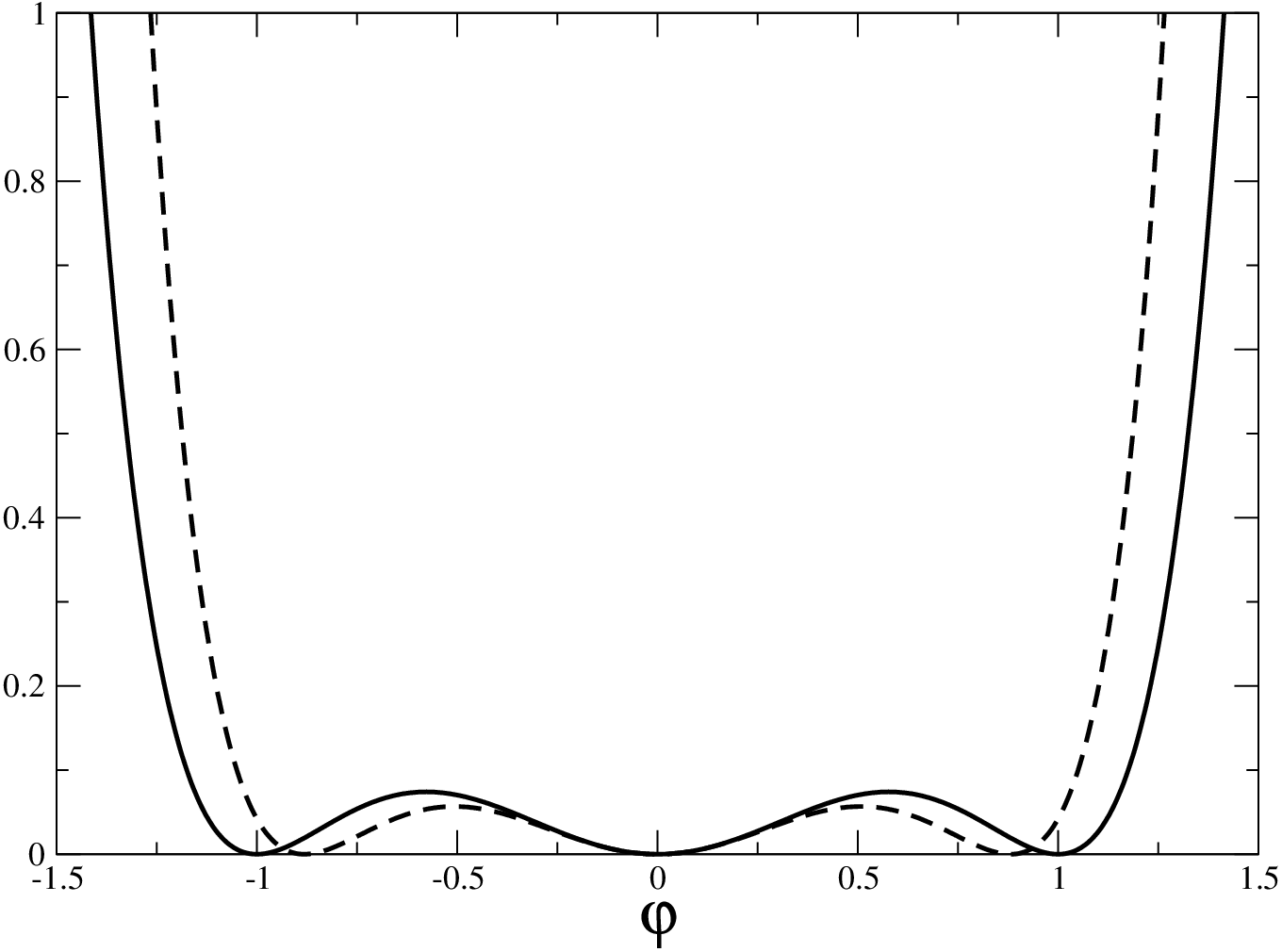}
		\label{fig:Ufieldphi6_a}}
	\quad 
	\subfigure[$a=0.5$]{
		\includegraphics[scale=0.22]{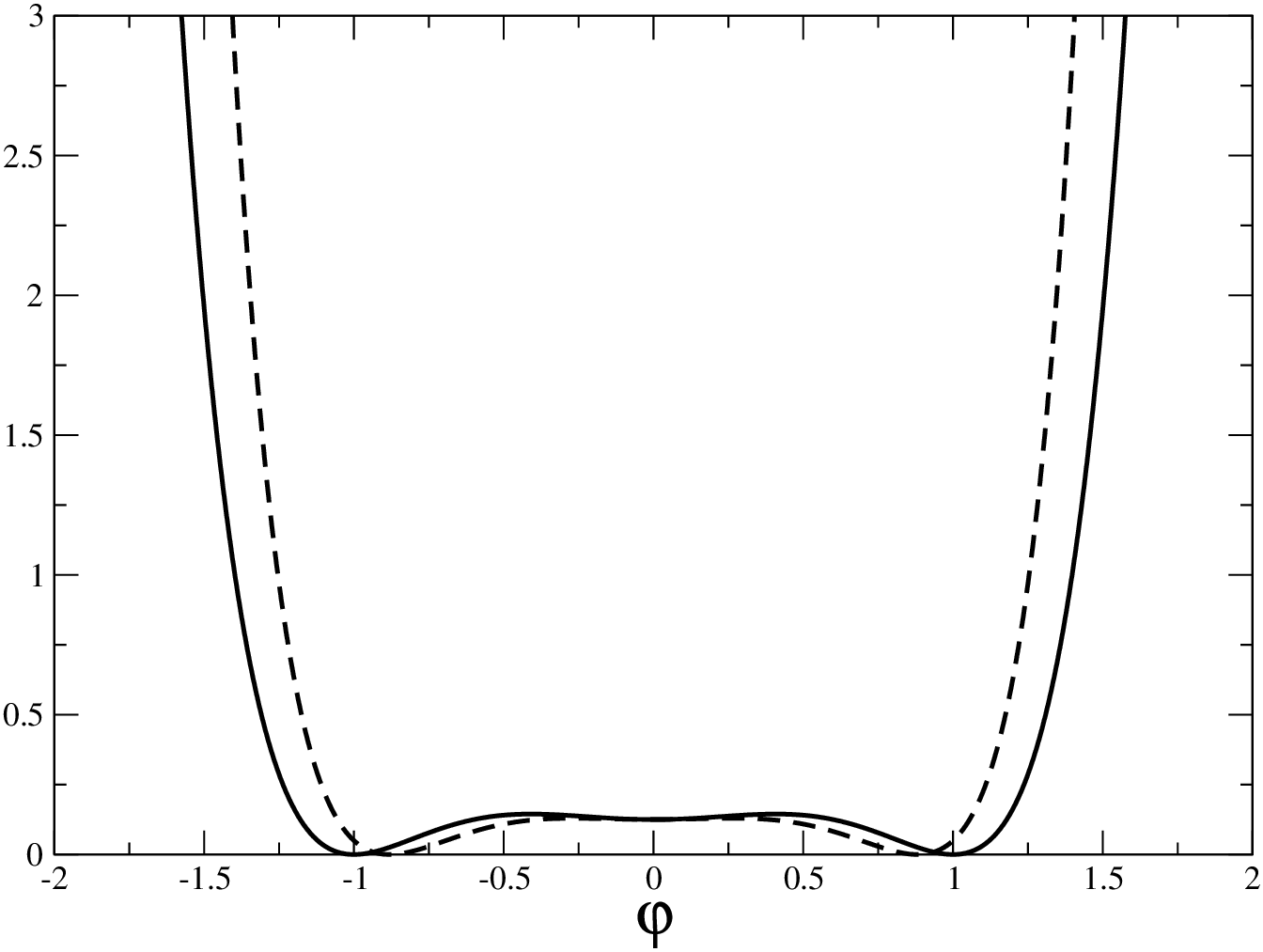}
		\label{fig:Ufieldphi6_b}}
	\quad 
	\subfigure[$a=1$]{ 
		\includegraphics[scale=0.22]{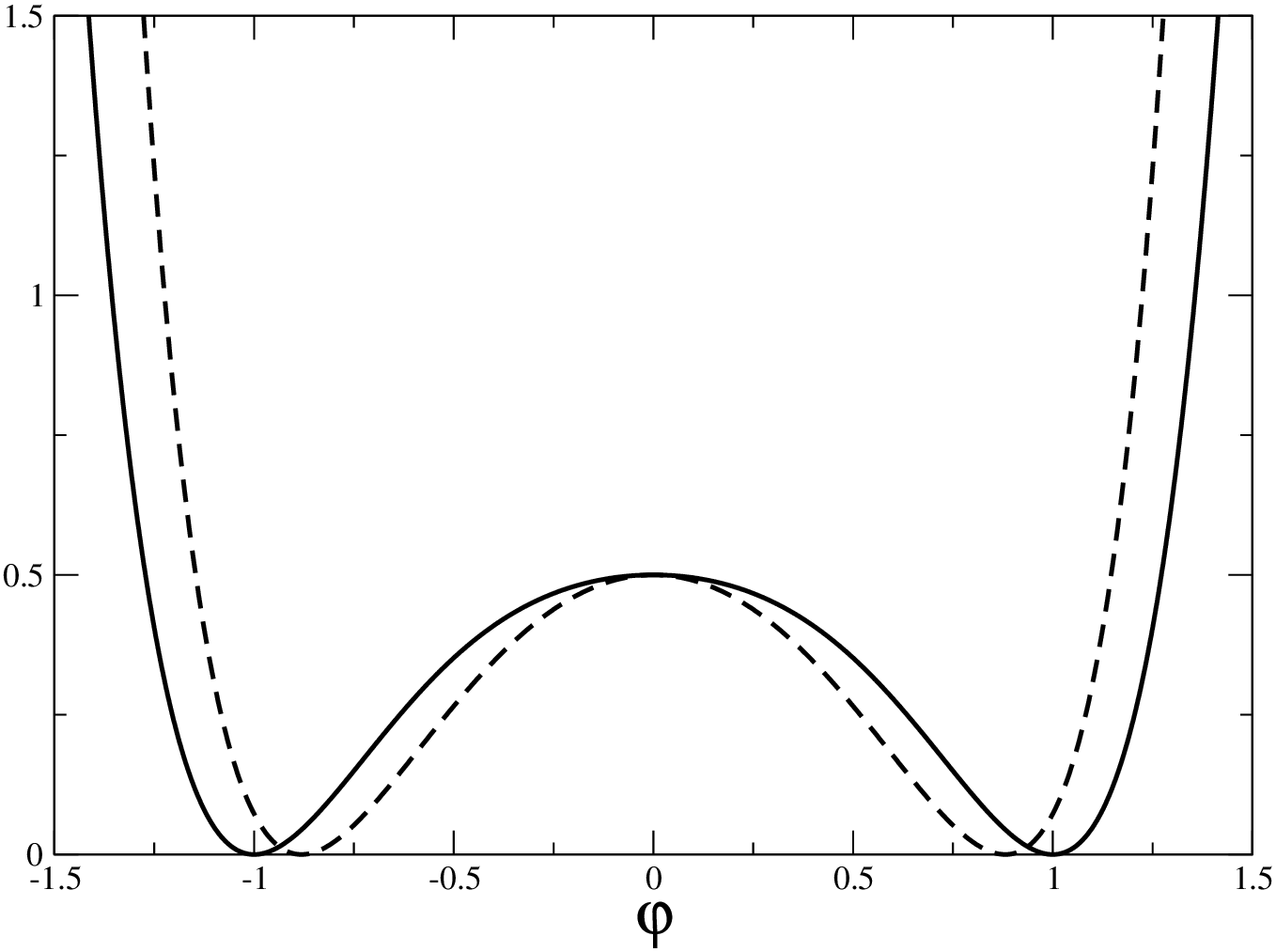}
		\label{fig:Ufieldphi6_c} }
	\caption{\label{fig:Ufieldphi6}The field potentials of the sinh-deformed $\varphi^6$ model (dashed lines) and polynomial $\varphi^6$ model (solid lines) for various values of the dimensionless parameter.}
\end{figure}

We consider the $\phi^{4}$ and $\varphi^{6}$ potentials
\begin{equation}
	U_{4}=\frac{1}{2}\left(\phi^{2}-1\right)^{2} \quad \text{and} \quad U_{6}=\frac{1}{2}\left(\varphi^{2}+a^{2}\right)\left(\varphi^{2}-1\right)^{2},
	\label{eq:models}
\end{equation}
where we have scaled all coordinates, fields and coupling constants such that only the $\varphi^{6}$ potential contains a single dimensionless parameter $a$. 
Applying the deformation procedure of Eq.(\ref{eq:deform_proc}) by using the deforming function $g(\phi)=\sinh \phi$ ($g(\varphi)=\sinh \varphi$ for the $\varphi^{6}$ model) we obtain the potentials of the sinh-deformed $\phi^{4}$ and $\varphi^{6}$ models, respectively as 
\begin{equation}
	V_{4}=\frac{1}{2}\sech^{2}\phi \left(1-\sinh^{2}\phi\right)^{2} \quad \text{and} \quad V_{6}=\frac{1}{2}\sech^{2}\varphi\left(\sinh^{2}\varphi+a^{2}\right)\left(1-\sinh^{2}\varphi\right)^{2}.
	\label{eq:sinhdefmod}
\end{equation}
Shown in Figure \ref{fig:kinkinterphimodel_a} are the field potential of the polynomial and sinh-deformed functions of the $\phi^{4}$ model. That for the $\varphi^{6}$ model for various values of $a$ are shown in Figure \ref{fig:Ufieldphi6}.
There are two vacuum solutions in the sinh-deformed $\phi^{4}$ model, $\phi_{0}=\pm {\rm arsinh} (1)$ as observed in Figure \ref{fig:kinkinterphimodel_a}, but for the sinh-deformed $\varphi^{6}$ model, three cases emerges. For $a=0$, we observe three degenerate minima at $\varphi_{0}=0$ and $\varphi_{0}=\pm {\rm arsinh} (1)$, see Figure \ref{fig:Ufieldphi6_a}. 

For $\displaystyle 0 < a^{2} \leq \frac{1}{2}$, we observed in Figure \ref{fig:Ufieldphi6_b} an additional local minimum at $\varphi_{0}=0$ and finally, for $\displaystyle a^{2} > \frac{1}{2} $ we observe in Figure \ref{fig:Ufieldphi6_c} two degenerate minima at $\varphi_{0}=\pm {\rm arsinh} (1)$. The models in Eq. (\ref{eq:sinhdefmod}) possesses a discrete symmetry $V_{4}(\phi)=V_{4}(-\phi)$ ($V_{6}(\varphi)=V_{6}(-\varphi)$) that is broken by the perturbative vacua $\phi_{0}=\pm   {\rm arsinh} (1)$ (respectively, $\varphi_{0}=0,\pm  {\rm arsinh} (1)$ in all three cases of the $\varphi^{6}$ sinh-deformed model). 

Applying Eq.(\ref{eq:defmakink}) to the kink-antikink solutions of the polynomial $\phi^{4}$ model \cite{Takyi:2016tnc} the kink-antikink solutions of the sinh-deformed $\phi^{4}$ model are
\begin{equation}
	\phi_{K,\overline{K}}(x)=\pm {\rm arsinh}(\tanh(x))\,,
	\label{eq:kink4}
\end{equation}
which are related by the spatial reflection $x \leftrightarrow -x$. Shown in Figure \ref{fig:kinkinterphimodel_b} are the kink solutions of the sinh-deformed and polynomial $\phi^{4}$ model. The corresponding classical kink mass is 
\begin{equation}
	E_{\rm cl} = \pi-2.
\end{equation}

\begin{figure}
	\centering
	\includegraphics[scale=0.3]{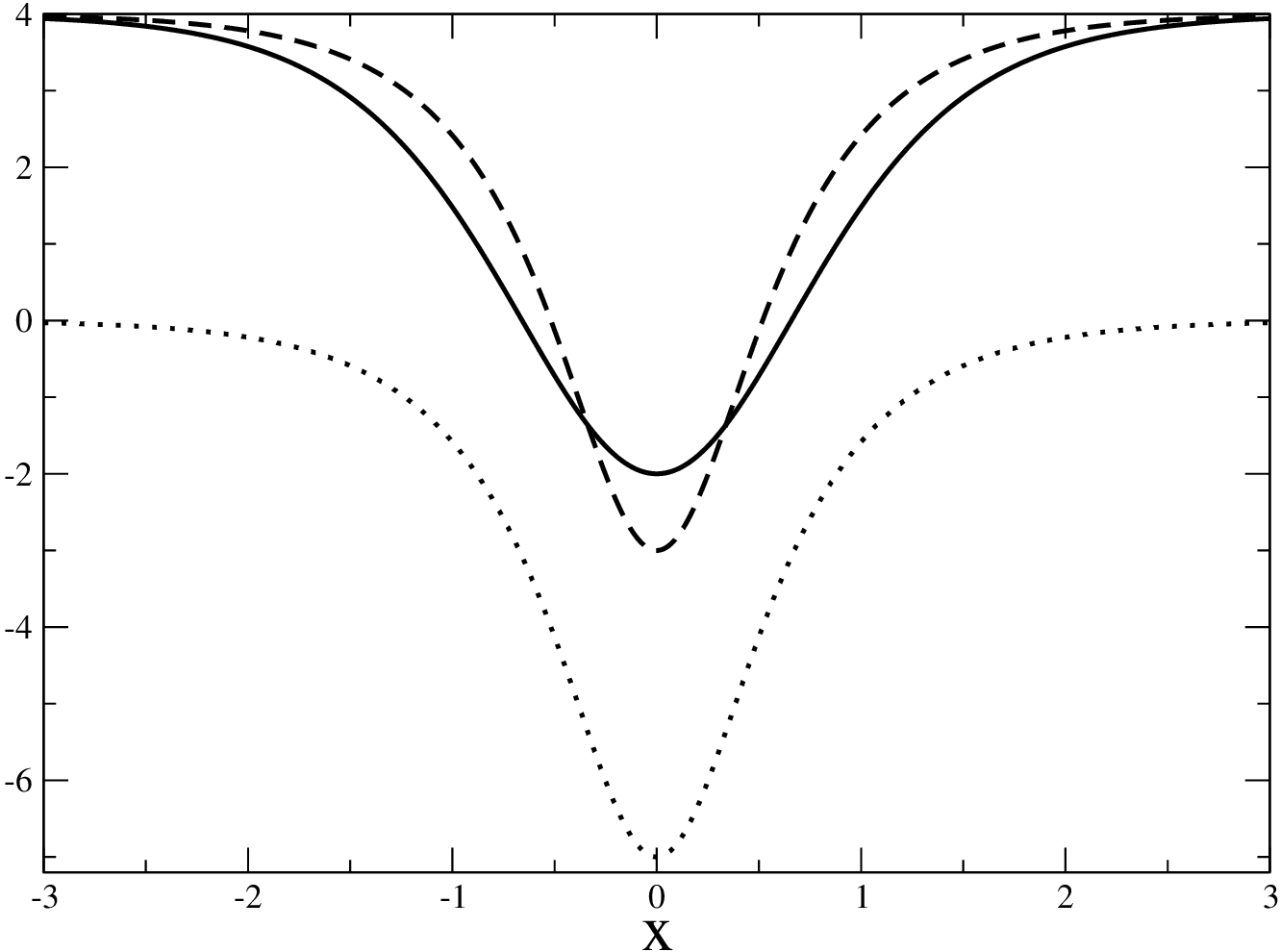}
	\caption{\label{fig:stabilitypotphi4}
		The scattering(dashed lines)-and pseudo(dotted lines)-potentials for the sinh-deformed $\phi^{4}$ model. The scattering potential for the corresponding polynomial model is indicated with the solid line. }
\end{figure}

The background potential for the fluctuations, which defines the excitation spectrum of the kink, is symmetric under the spatial reflection $x \rightarrow -x$ 
\begin{equation}
	v_{4}(x)=2\tanh^{2}x + 1 + \frac{8\tanh^{2}x -4}{\left(1+\tanh^{2}x\right)^{2}}.
\end{equation}
This yield $m_{R}=m_{L}$, with $m_{R}^{2}=4$ as observed in Figure \ref{fig:stabilitypotphi4}.

\begin{figure}
	\centering
	\subfigure[Kink solutions cf. eq(\ref{eq:kinkphi6}) for $a=1$ ]{\includegraphics[scale=0.22]{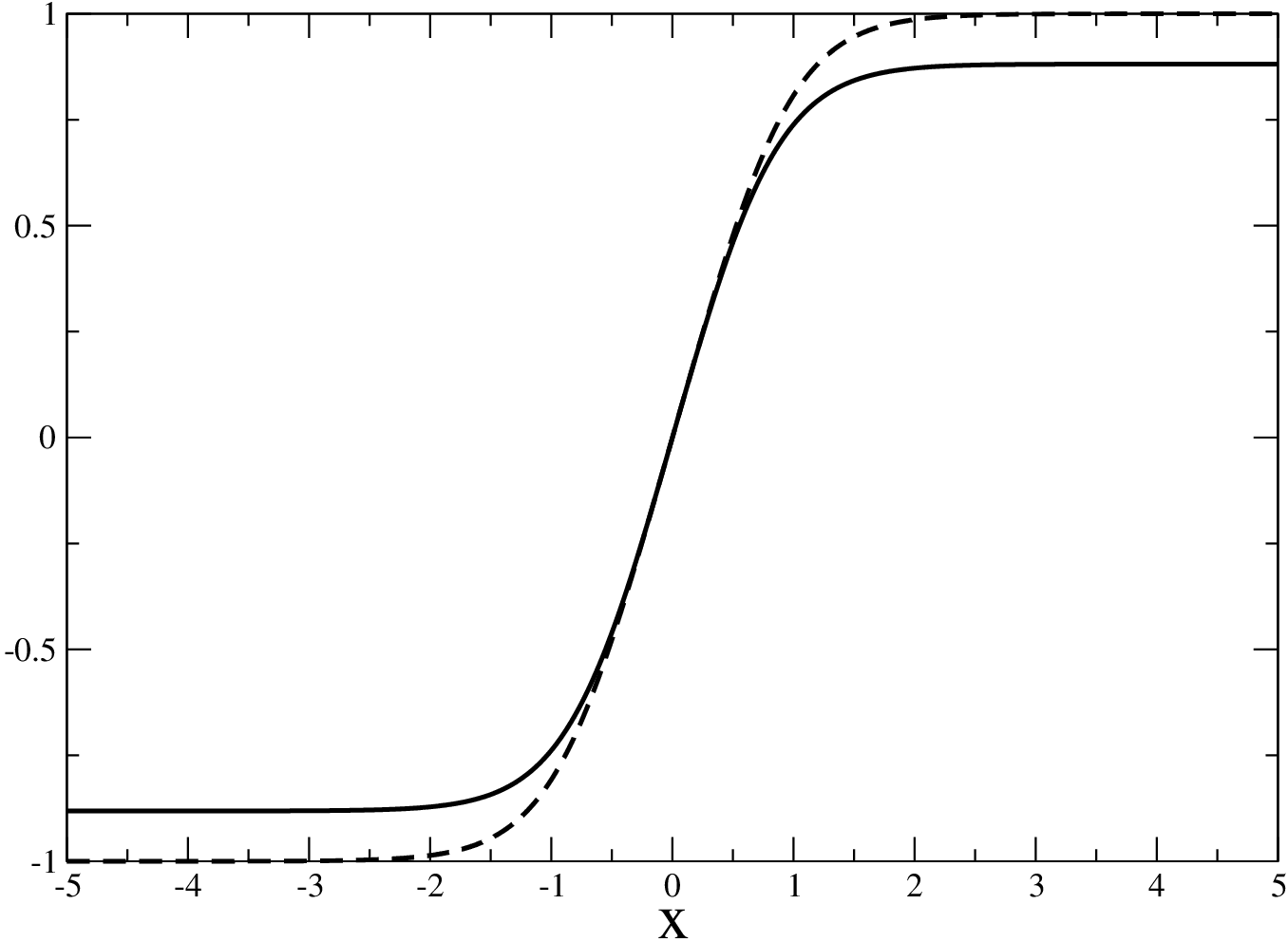}
		\label{fig:kinkphi6_a}}
	\quad 
	\subfigure[ Kink solutions cf. eq(\ref{eq:kinkIphi6}) for $a=0$]{
		\includegraphics[scale=0.22]{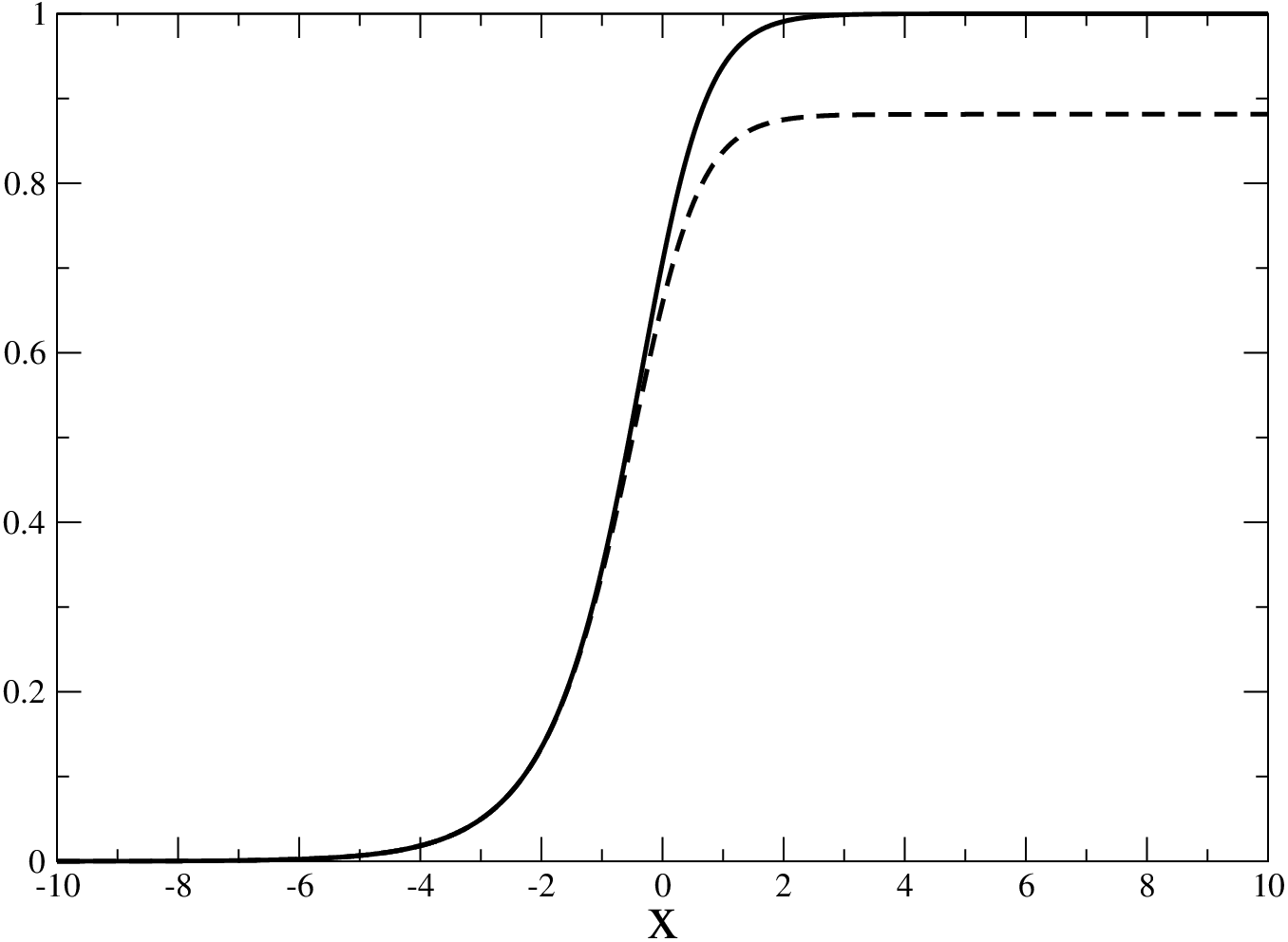}
		\label{fig:kinkphi6_b}}
	\quad 
	\subfigure[Kink solutions cf. eq(\ref{eq:kinkIIphi6}) for $a=0$]{ 
		\includegraphics[scale=0.22]{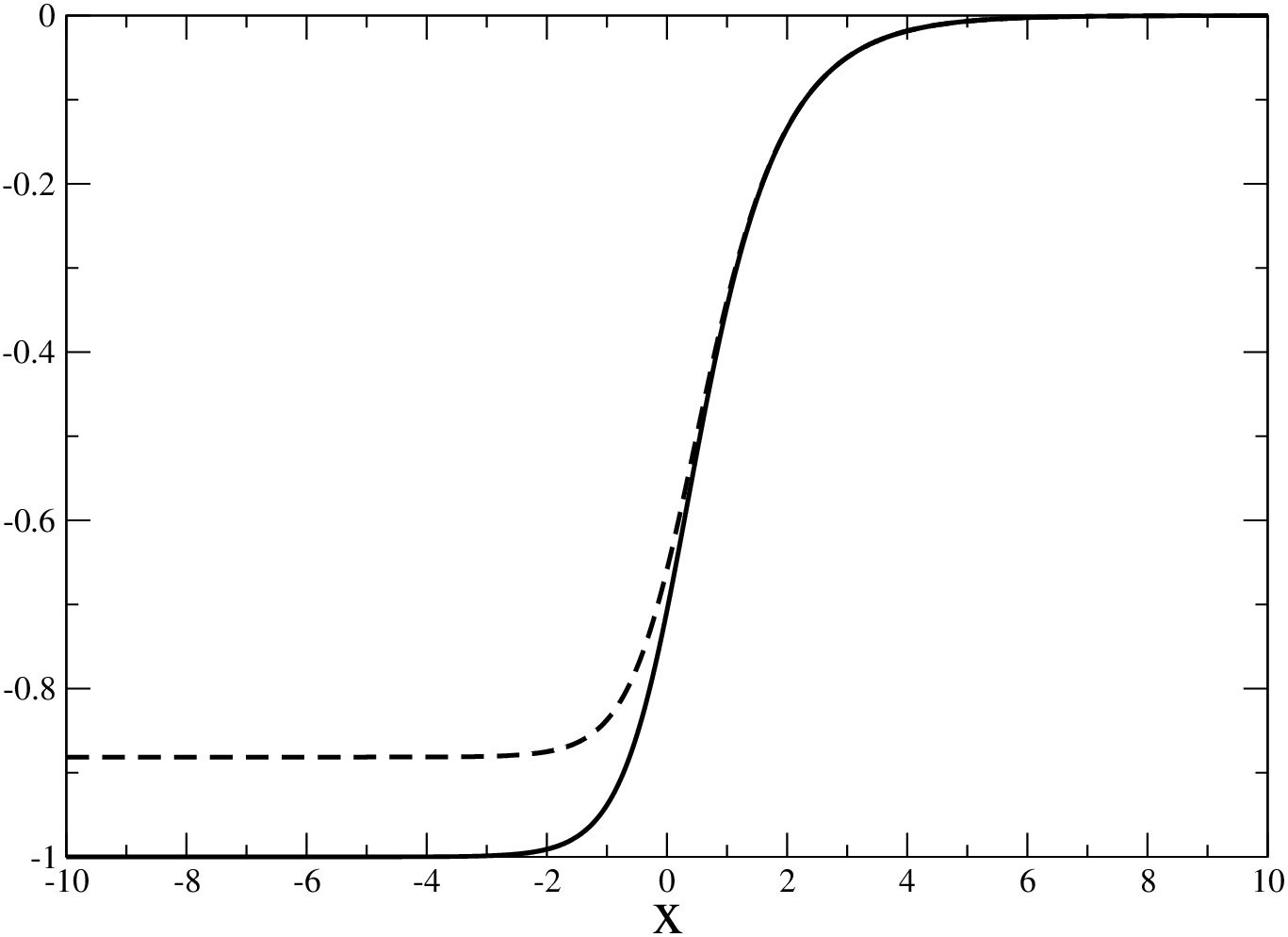}
		\label{fig:kinkphi6_c} }
	\caption{\label{fig:kinkphi6}The kink solutions of the sinh-deformed $\varphi^6$ model (dashed lines) and polynomial $\varphi^6$ model (solid lines).}
\end{figure}

\begin{figure}
	\centering
	\subfigure[The sinh-deformed and polynomial $\varphi^{6}$ model for $a=1$]{
		\includegraphics[scale=0.3]{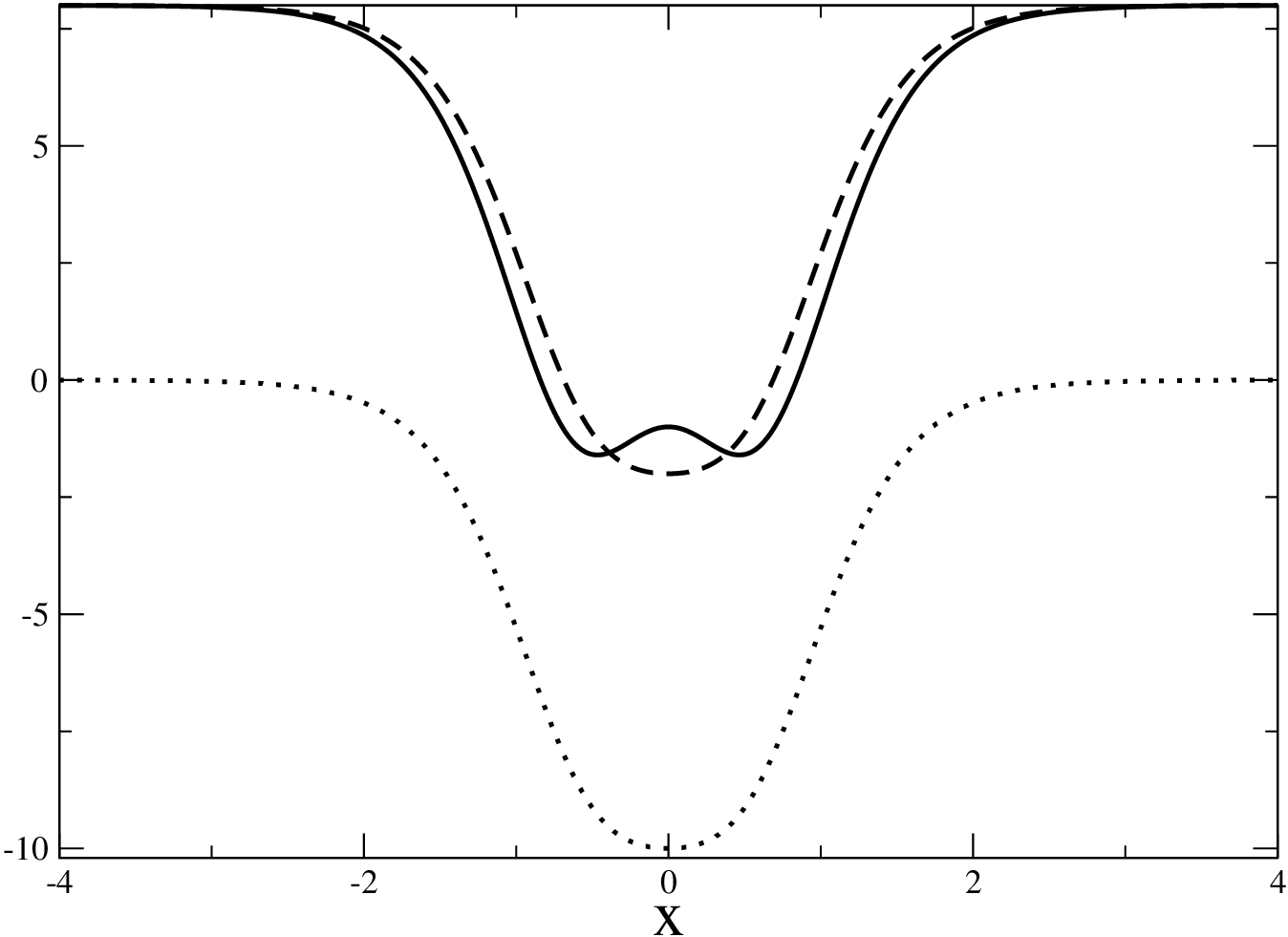}
		\label{fig:stabilitypot_a}}
	\quad 
	\subfigure[The sinh-deformed and polynomial $\varphi^{6}$ model for $a=0$ ]{\includegraphics[scale=0.3]{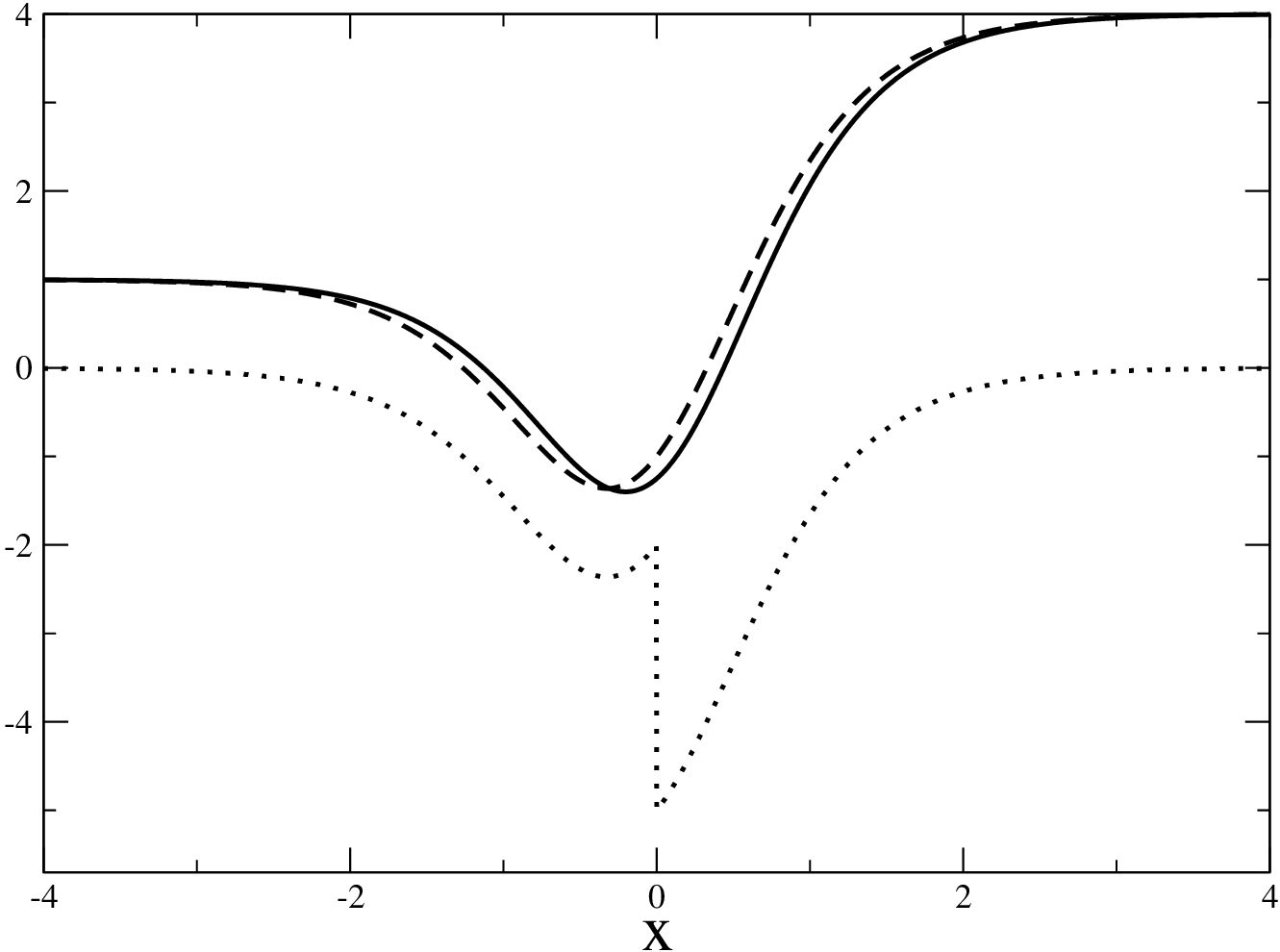}
		\label{fig:stabilitypot_b}}
	\caption{\label{fig:stabilitypot6}
		The scattering(dashed lines)- and pseudo(dotted lines)-potentials for the sinh-deformed $\varphi^{6}$ model. The scattering potential for the corresponding polynomial model is indicated with the solid line. }
\end{figure}

In the sinh-deformed $\varphi^{6}$ model, we obtain the kink solution for $a\neq 0$ as \cite{Weigel:2017iup,Lohe:1979mh}
\begin{equation}
	\varphi_{K}={\rm arsinh} \left[a \frac{X-1}{\sqrt{4X+a^{2}(1+X)^{2}}}\right], \quad \text{with} \quad X=e^{m_{R}x}
	\label{eq:kinkphi6}
\end{equation}
where $m_{R}=2\sqrt{1+a^{2}}$. The kink solution for $a=1$ is indicated in Figure \ref{fig:kinkphi6_a}. The spatial reflection $x \rightarrow -x$ gives the antikink. The background potential is
\begin{align}
	v_{6}(x) & =2(a^{2}+\sinh^{2}\varphi_{K})(3\sinh^{2}\varphi_{K}-1)+ (\sinh^{2}\varphi_{K}-1)(9\sinh^{2}\varphi_{K}-1) \nonumber \\
	& -3(\sinh^{2}\varphi_{K}-1)^{2}\tanh^{2}\varphi_{K} - 6 (a^{2}+\sinh^{2}\varphi_{K})(\sinh^{2}\varphi_{K}-1)\tanh^{2}\varphi_{K} \nonumber \\
	&+ (a^{2}+\sinh^{2}\varphi_{K})(\sinh^{2}\varphi_{K}-1)^{2}(2\sinh^{2}\varphi_{K}-1) {\rm sech}^{4}\varphi_{K}
\end{align}
and is symmetric under the spatial reflection $x \rightarrow -x$ as observed in Figure \ref{fig:stabilitypot_a} for $a=1$. Consequently, $m_{L}=m_{R}$. 

For $a=0$, two distinct kink solutions exist \cite{Takyi:2016tnc,Dorey:2011yw}. The first one interpolates between $\varphi_{0}=0$ and $\varphi_{0}={\rm arsinh}(1)$
\begin{equation}
	\varphi_{K_{I}} = {\rm arsinh} \left(\sqrt{\frac{1}{2}\left(1 + \tanh x\right)}\right),
	\label{eq:kinkIphi6}
\end{equation} 
which we show in Figure \ref{fig:kinkphi6_b}, while the second one interpolates between $\varphi_{0}=-{\rm arsinh}(1)$ and $\varphi_{0}=0$ (see Figure \ref{fig:kinkphi6_c}),
\begin{equation}
	\varphi_{K_{II}} = -{\rm arsinh} \left(\sqrt{\frac{1}{2}\left(1 - \tanh x\right)}\right).
	\label{eq:kinkIIphi6}
\end{equation} 
The corresponding classical kink mass in either case is
\begin{equation}
	E_{\rm cl}=\frac{1}{2}\left(2 \ln 2 -1\right).
\end{equation}
The background potential for the fluctuations in this case is not symmetric under the spatial reflection $x \rightarrow -x$
\begin{equation}
	v_{6}(x) = 2 \tanh^{2} x + 5 \tanh x - 7 + \frac{10 \tanh^{2}x + 54}{\left(3+ \tanh^{2}x\right)^{2}}.
\end{equation}
This, of course, implies that $m_{R}=2 \neq m_{L}=1$ as observed in Figure \ref{fig:stabilitypot_b}. In the same figure we also show the corresponding pseudo-potential.

\subsection{The Hyperbolic Model}
\begin{figure}
	\centering
	\subfigure[Field potential of the hyperbolic (dashed line) and polynomial (solid line) $\phi^{4}$ model.]{
		\includegraphics[scale=0.25]{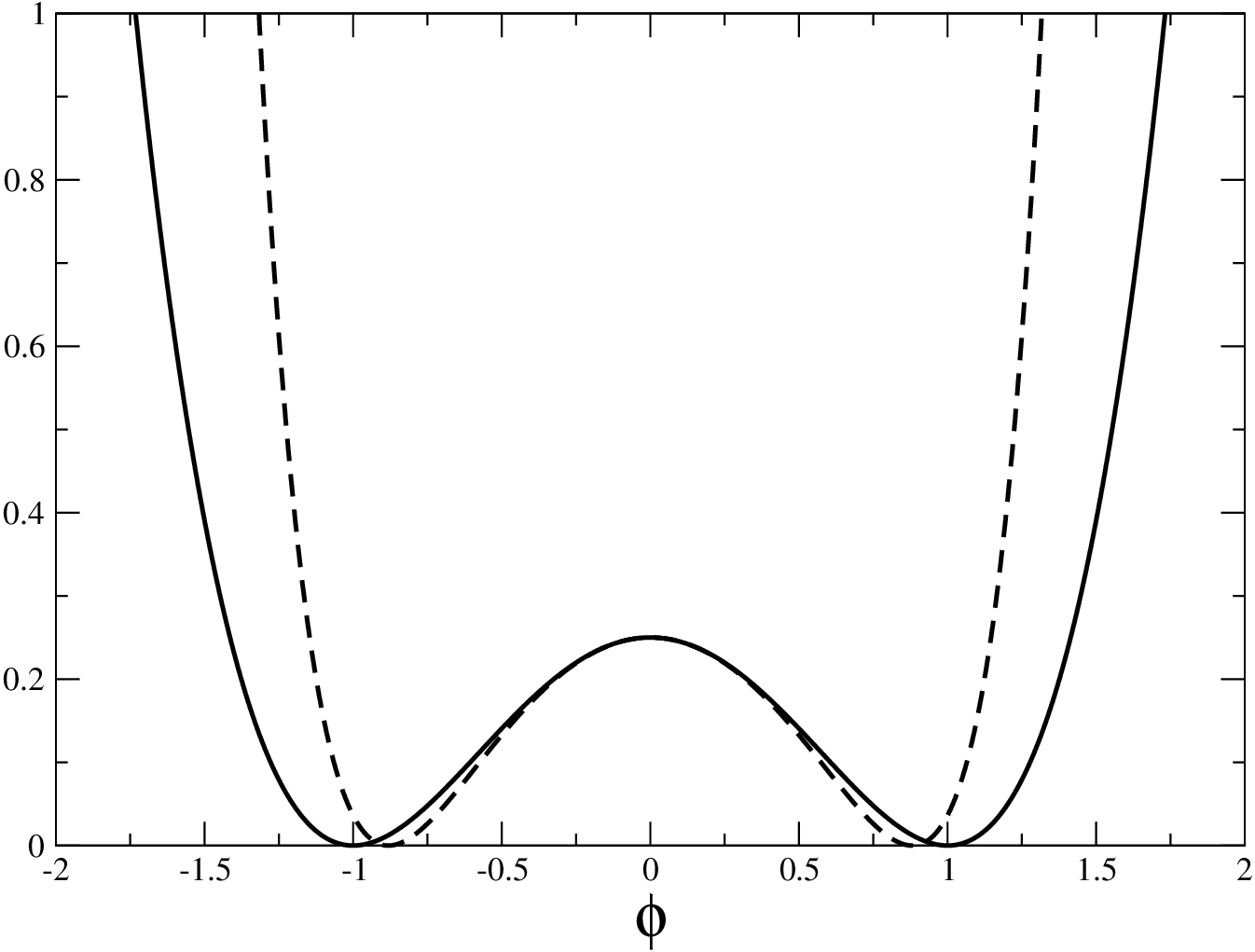}
		\label{fig:kinkinterhyper_a}}
	\quad 
	\subfigure[Kink solutions of the hyperbolic (dashed line) and polynomial (solid line) $\phi^{4}$ model.]{
		\includegraphics[scale=0.25]{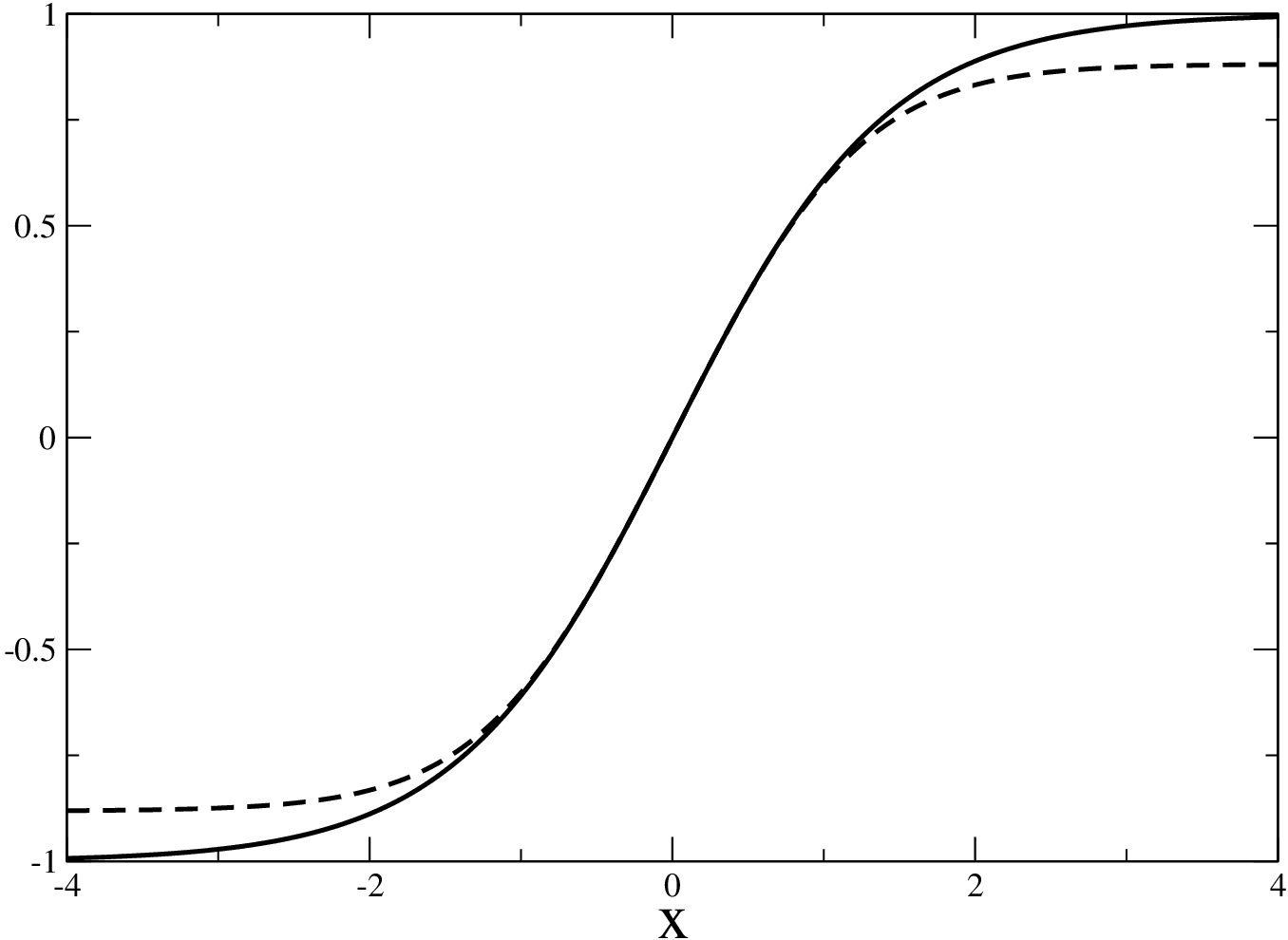}
		\label{fig:kinkinterhyper_b} }
	\caption{\label{fig:kinkinterhyper} The field potential and kink solutions of the hyperbolic $\phi^{4}$ model.}
\end{figure}

\begin{figure}
	\centering
	\includegraphics[scale=0.3]{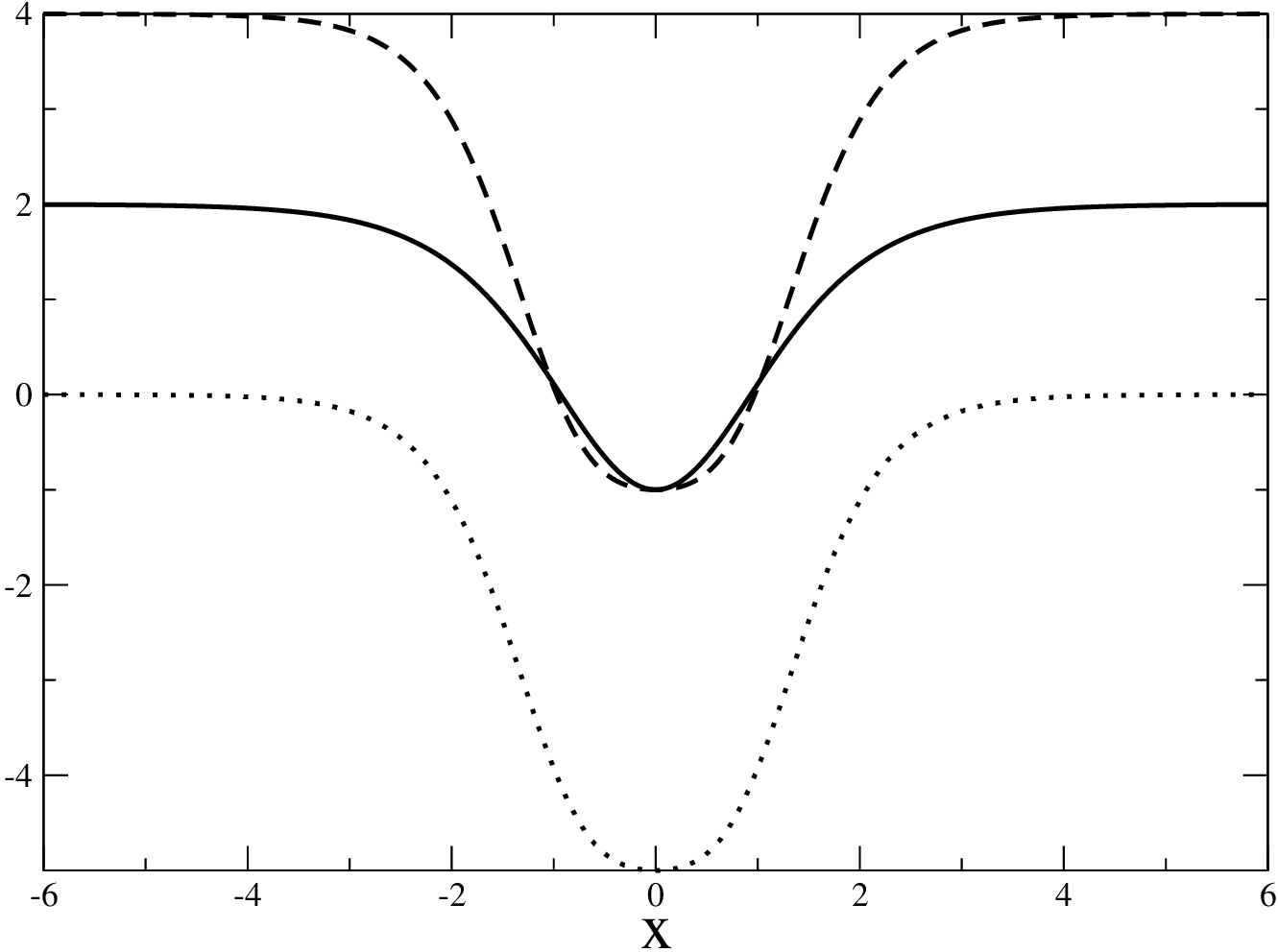}
	\caption{\label{fig:stabilitypothy}
		The scattering(dashed lines)-and pseudo(dotted lines)-potentials for the hyperbolic $\phi^{4}$ model. The scattering potential for the corresponding polynomial model is indicated with the solid line. }
\end{figure}

Here we consider the hyperbolic potential \cite{Bazeia:2018qur}
\begin{equation}
	V(\phi)=\frac{1}{4} \left(1-\sinh^{2}(\phi)\right)^{2}.
	\label{eq:hyper4}
\end{equation}
This potential has two degenerate minima $\phi_{0}=\pm {\rm arsinh}(1)$ as observed in Figure \ref{fig:kinkinterhyper_a}. In the limit of small values of $\phi$, this potential approaches the polynomial $\phi^{4}$ model 
\begin{equation}
	U(\phi)=\frac{1}{4}\left(1-\phi^{2}\right)^{2},
	\label{eq:phi4fhalve}
\end{equation}
with a factor $\frac{1}{2}$ compared to $U_{4}$ in Eq. (\ref{eq:models}). The kink-antikink solutions of the hyperbolic $\phi^{4}$ model are
\begin{equation}
	\phi_{K,\overline{K}}(x)=\pm {\rm artanh}(\frac{1}{\sqrt{2}}\tanh(x))\,,
	\label{eq:kinkheper4}
\end{equation}
which are related by the spatial reflection $x \leftrightarrow -x$. The kink solution is shown in Figure \ref{fig:kinkinterhyper_b}. The resulting classical mass is 
\begin{equation}
	E_{\rm cl}=\frac{3}{8}\sqrt{2} \Big\{ \ln\left(2\sqrt{2}+3\right) - \ln\left(-2\sqrt{2}+3\right)\Big\}-1.
\end{equation}
The background potential 
\begin{equation}
	v(x)=2 + \frac{14 \tanh^{2}(x) - 12}{\left(\tanh^{2}(x)-2\right)^{2}}
\end{equation}
is symmetric under the spatial reflection $x \leftrightarrow -x$ as observed in Figure \ref{fig:stabilitypothy}. In the limit as $x \rightarrow \pm \infty$ we have $m_{L}=m_{R}=2$. 

\section{Numerical Results}
\label{results}
In this section, we report our numerical results for the VPEs for the two models discussed above. As stated earlier, we have rescaled to dimensionless coordinates and fields such that model parameters (coupling constant `$\lambda$' and mass `$m$') are unity. It must be noted that the coupling constant $\lambda$ serves as a loop-counting parameter. In this case, the classical mass scales as ~$\frac{1}{\lambda}$ while the VPE proportional to $m$, where $m$ is a mass parameter \footnote{This generally holds for the $\phi^{4}$ model. For this to be correct, for the $\varphi^{6}$ model the dimensionless parameter $a$ must be written as $a=\alpha\sqrt{\frac{m}{\lambda}}$ where $\alpha$ do not vary with $m$ or $\lambda$. Then the quadratic mass type term in $U_{6}(\varphi)$ does not contain the coupling constant $\lambda$.} in the potentials considered. Thus, taking the coupling constant as unity does not affect the computation of the VPE.

We obtain the bound states energies contribution to the VPE, by first solving the Schr\"odinger-wave equation using the Boole's algorithm, which is otherwise know as the $5$-point closed Newton-Cote formula; with the initial conditions
\begin{equation*}
	\eta_{R} \rightarrow 1, \quad \eta_{R}^{\prime} \rightarrow - \sqrt{m_{R}^{2}-m_{L}^{2}-E^{2}} \quad \text{as} \quad x \rightarrow \infty
\end{equation*}
and 
\begin{equation*}
	\eta_{L} \rightarrow 1, \quad \eta_{L}^{\prime} \rightarrow \sqrt{m_{L}^{2}-E^{2}} \quad \text{as} \quad x \rightarrow -\infty.
\end{equation*}
from either side. We obtained the bound state energy by tuning the energy $\omega=\omega_{j}$ such that the Wronskian $\eta_{L} \eta_{R}^{\prime} - \eta_{R}\eta_{L}^{\prime}$ is zero at the matching point $x_{m}$. 

The next step requires computing the continuum part of the VPE, where the phase shift and its first Born approximation are obtained by solving Eqs. (\ref{eq:phaseshift}) and (\ref{eq:Born}). We achieved this by utilizing the Runge Kunta algorithm to solve Eq. (\ref{eq:parametequa}) which yields the S-matrix needed for the computation of the phase shift. It must be noted that the numerically obtained phase shift does not vary with the choice of $x_{m}$. 

We will be looking at two scenarios in the computation of the VPE. The first case deals with models with symmetric scattering potential. This regime, has equivalent vacua at $x \rightarrow \pm \infty$, as such translating the kink solution as\footnote{This corresponds to kink solution centered at $-x_{0}$.} $x \rightarrow x+x_{0}$ does not change the VPE of the kink. For models with asymmetric scattering potential, the non-equivalent vacua produce translational variance with the center of the kink solution $x_{0}$.  

\subsection{The Sinh-Deformed Models}
We first compare the results of the sinh-deformed $\phi^{4}$ model and its corresponding polynomial model. The two vacua for both models have equal curvature with $m_{L}=2$. From Figure \ref{fig:stabilitypotphi4}, one observes that the scattering potential of the polynomial $\phi^{4}$ model is shallower and broader than that of the polynomial $\phi^{4}$ model. We observed that both cases have two bound states with the bound state energies of the polynomial $\phi^{4}$ model smaller than that of the corresponding sinh-deformed model; this is attributed to the shallowness and broadness of its scattering potential. This in effect accounts for the smaller value of the VPE observed in the polynomial $\phi^{4}$ model as indicated in Table  \ref{t1}. Figure \ref{fig:phase_a} shows the phase shift for the sinh-deformed $\phi^{4}$ model, confirming the Levinson's theorem cf. Eq. \ref{eq:levisonthe} for $n=2$.
\begin{table}
	\centerline{
		\begin{tabular}{c |c c|c|c|c}
			&\multicolumn{2}{c|}{bound state energies} 
			& $E_{\rm b.s}$ & $E_{\rm scat.}$ & $E_{\rm vac.}$\cr
			\hline
			polynomial $\phi^{4}$    & 0.0 & 1.732  & -1.134 & 0.470 & -0.664(-0.666) \cr
			\hline
			sinh-deformed $\phi^{4}$ & 0.0 & 1.892  & -1.054 & 0.412 & -0.643(-0.644) \cr
	\end{tabular}}
	\caption{\label{t1}The bound state energies and VPEs of the sinh-deformed $\phi^{4}$ model cf. Eq. (\ref{eq:sinhdefmod}) and polynomial $\phi^{4}$ model cf. Eq. (\ref{eq:models}). The entries $E_{\rm b.s}$ and $E_{\rm scat.}$ denote the bound state and continuum
		contributions to the VPE, {\it i.e.} the two distinct terms in Eq.~(\ref{eq:evac}). The last entry in parenthesis are the Jost solutions cf. Eq.~(\ref{eq:Jost}) confirming the VPE result.}
\end{table}
\begin{table} 
	\caption{\label{t2} The VPE as a function of the center of the kink $x_0$ in the polynomial $\varphi^{6}$ model and sinh-deformed $\varphi^{6}$  model cf. Eq. (\ref{eq:sinhdefmod}) for $a=0$.}
	\centerline{
		\begin{tabular}{c| c c c c c  }
			&\multicolumn{5}{c}{$E_{vac}$} \cr 
			\hline 
			$x_{0}$ & $-2$  & $-1$  & $0$  & $1$  & $2$   \cr \hline 
			polynomial $\varphi^{6}$ & 0.154 & 0.053 & -0.047 & -0.148 & -0.249 \cr 
			sinh-deformed $\varphi^{6}$ & 0.229 & 0.129 & 0.028 & -0.074 & -0176
		\end{tabular}
	}
\end{table} 

\begin{table} 
	\caption{\label{t3} Comparison of the VPEs for the symmetric scattering potential of the polynomial $\varphi^{6}$ model cf. Eq. (\ref{eq:models}) and sinh-deformed $\varphi^{6}$ model cf. Eq.(\ref{eq:sinhdefmod}) for $a\neq 0$. We have also indicated the Jost solutions for both the polynomial and the sinh-deformed $\varphi^{6}$ models. The error is calculated as the relative error $\frac{\vert E_{{\rm vac}} -E_{{\rm vac}}^{S} \vert }{\vert E_{{\rm vac}} \vert}$. Where $E_{{\rm vac}}$ is the method of computing the VPE using Eq. (\ref{eq:evac}) and $E_{{\rm vac}}^{S}$ is the method of computing the VPE using the Jost function formalism of Eq. (\ref{eq:Jost}).}
	\centerline{
		\begin{tabular}{c| c c c c c c }
			$a$                         & $0.01$  & $0.05$  & $0.1$  & $0.2$  & $1.0$  & $1.5$   \cr \hline 
			polynomial $\varphi^{6}$    & $-1.841$  & $-1.596$  & $-1.462$ & $-1.297$ & $-1.102$ & $-1.297$ \cr 
			Jost (poly. $\varphi^{6}$)  & $-1.840$  & $-1.595$  & $-1.461$ & $-1.298$ & $-1.101$ & $-1.295$ \cr 
			Error (poly. $\varphi^{6}$) & $3.04\times 10^{-4}$  & $6.84\times 10^{-4}$& $6.41\times 10^{-4}$ & $8.52 \times 10^{-4}$ & $1.23 \times 10^{-3}$ & $1.30 \times 10^{-3}$ \cr \hline  
			sinh-deformed $\varphi^{6}$ & $-1.819$  & $-1.570$  & $-1.430$ & $-1.256$ & $-1.037$ & $-1.230$  \cr 
			Jost (sinh-def. $\varphi^{6}$) & $-1.827$  & $-1.576$  & $-1.435$ & $-1.261$ & $-1.038$ & $-1.229$ \cr 
			Error (sinh-def. $\varphi^{6}$) & $4.22\times 10^{-3}$  & $4.04\times 10^{-3}$& $3.72\times 10^{-3}$ & $3.33 \times 10^{-3}$ & $6.93 \times 10^{-4}$ & $5.44 \times 10^{-4}$ \cr \hline  
		\end{tabular}
	}
\end{table} 
In the case of sinh-deformed $\varphi^{6}$ model for $a=0$, we have an asymmetric scattering potential with an unequal meson masses $m_{L}=1$ and $m_{R}=2$ for the kink soliton cf. Eq. (\ref{eq:kinkIphi6}). We observed from Figure \ref{fig:stabilitypot_b} for $a=0$ that the scattering potential is narrower and almost overlap that of its polynomial counterpart. This in turn causes its VPE to be large cf. Table \ref{t2}. Furthermore, the results indicate that except for the zero-mode there will be no additional bound state. We numerically obtained the binding energy (denoted by $E_{\rm b.s}$), as $E_{\rm b.s}=-0.4107$ which is larger than the binding energy, $E_{\rm b.s}=-0.5$ of the polynomial $\varphi^{6}$ model. The unequal meson masses produces a translational variance of the VPE as seen in Table \ref{t2} under the translation $x \rightarrow x+x_{0}$. The results show that as $x_{0}$ increases, the kink shifts the vacuum with the bigger mass towards negative infinity causing low-lying modes to disappear. This decreases the VPE. It must be noted that the results stem from a change in values of the phase shift as $x_{0}$ varies. The phase shift exhibits the threshold scattering cusp $k=\sqrt{m_{R}^{2}-m_{L}^{2}}=\sqrt{3}$ and approaches $\frac{\pi}{2}$ as $k \rightarrow 0$ confirming Levinson's theorem. We verified this by the plot of the phase shift in Figure \ref{fig:phase_d}.

For $a\neq 0$ we observe a symmetric scattering potential cf. of Figure \ref{fig:stabilitypot_a}. This potential is narrower compared to that of the polynomial $\varphi^{6}$ model. In effect, it yields large values for the VPEs for various values of $a$ as seen in Table \ref{t3}. The results in Table \ref{t3} show that the results using the Jost function formalism of Eq. (\ref{eq:Jost}) agrees favorable well with the present calculations of Eq. (\ref{eq:evac}) with a minor error. The error margin improves as $a$ increases. Here the error is estimated as the relative error $\displaystyle \frac{\vert E_{{\rm vac}} -E_{{\rm vac}}^{S} \vert }{\vert E_{{\rm vac}} \vert}$. Also, we show in Figure \ref{fig:phase_c} the phase shift for $a=1$.

\subsection{The Hyperbolic Model}
\begin{figure}
	\centering
	\subfigure[Phase shift of the sinh-deformed $\phi^{4}$ model.]{
		\includegraphics[scale=0.27]{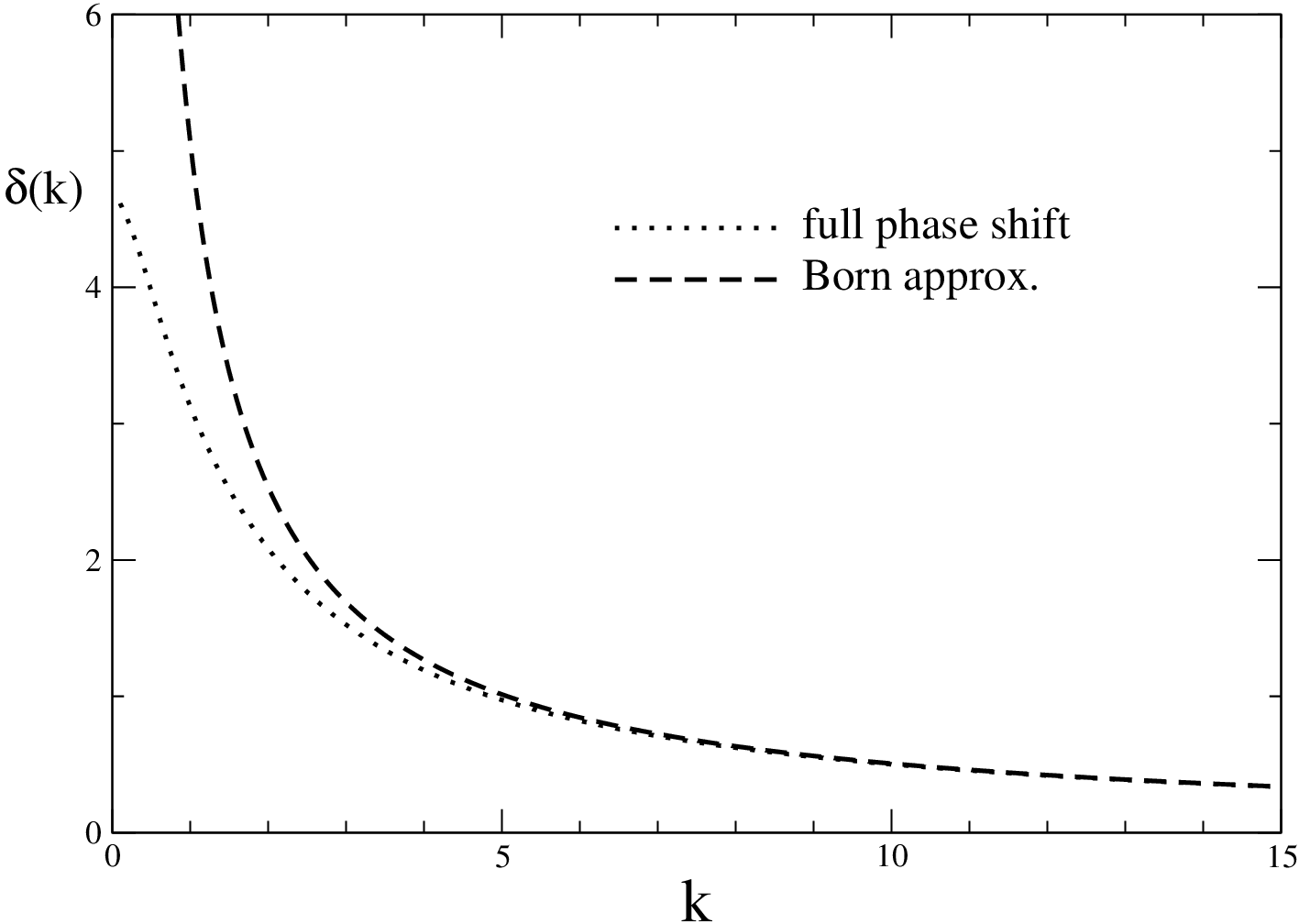}
		\label{fig:phase_a}}
	\quad 
	\subfigure[Phase shift of the hyperbolic $\phi^{4}$ model.]{
		\includegraphics[scale=0.27]{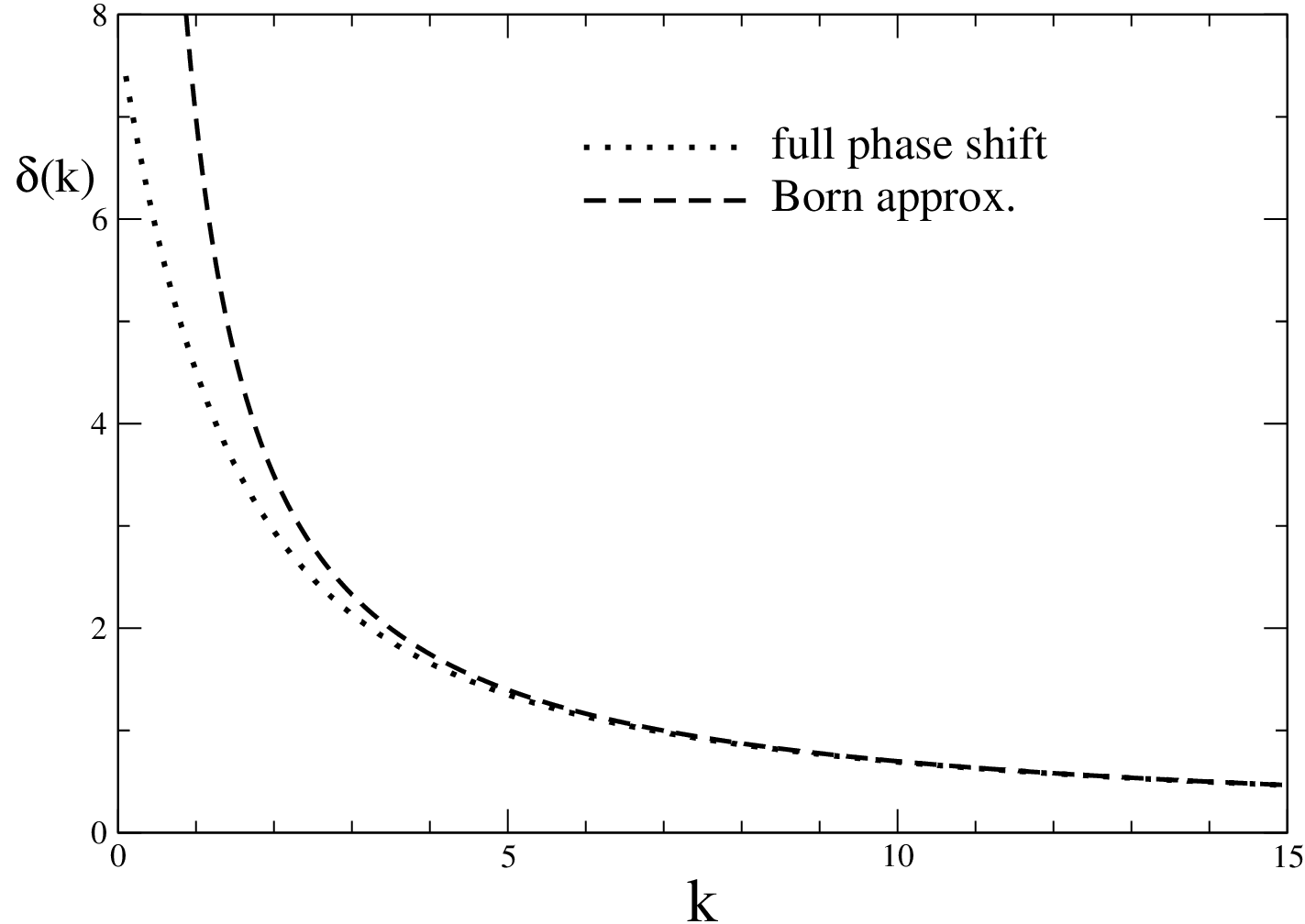}
		\label{fig:phase_b}}
	\quad
	\subfigure[Phase shift of the sinh-deformed $\varphi^{6}$ models for $a=1$.]{
		\includegraphics[scale=0.27]{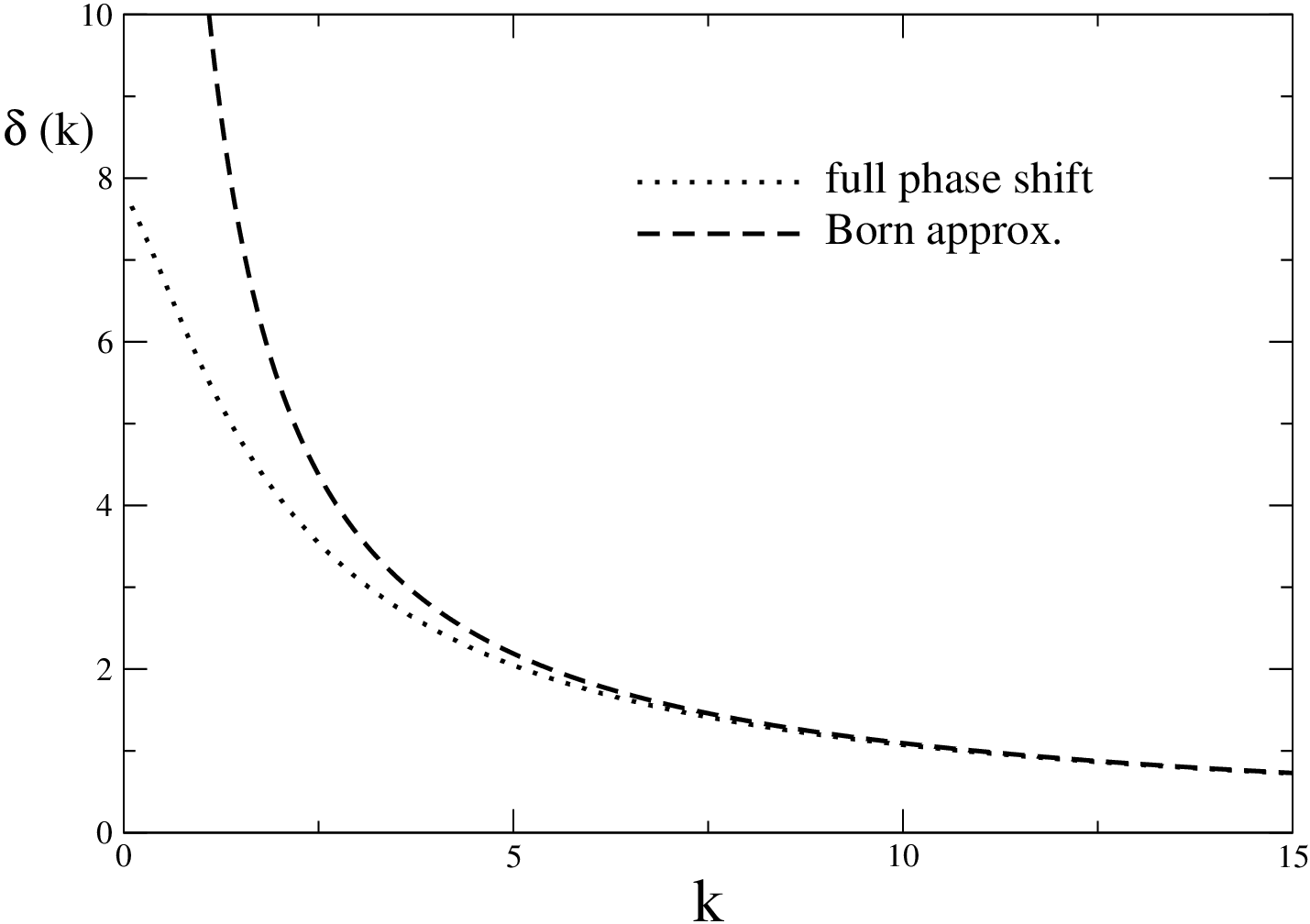}
		\label{fig:phase_c}}
	\quad
	\subfigure[Phase shift of the sinh-deformed $\varphi^{6}$ models for $a=0$.]{
		\includegraphics[scale=0.27]{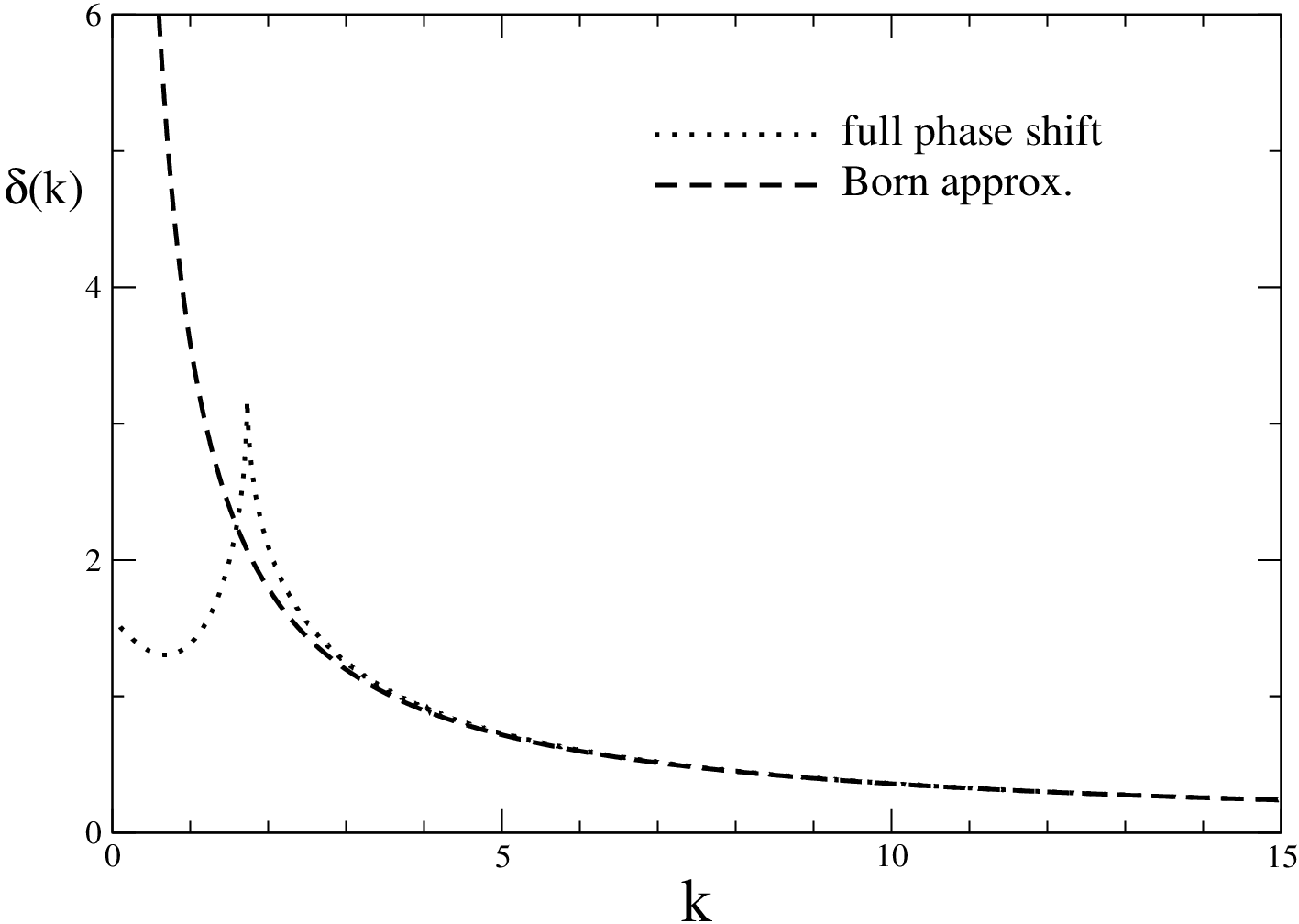}
		\label{fig:phase_d}}
	\caption{\label{fig:phase}Phase shifts for the various models.}
\end{figure}
We observe that the scattering potential of the hyperbolic $\phi^{4}$ model is symmetric and broader than its corresponding polynomial model cf. Figure \ref{fig:stabilitypothy}. This causes its VPE to be smaller compared to the polynomial counterpart as seen in Table \ref{t4}. Also, we observe two vibrational modes in addition to the translational zero-mode compared to the one vibrational mode of its polynomial model. The occurrence of this is due to the broadness of its scattering potential. Also, the VPE result from calculations of the hyperbolic model is in agreement (within numerical precision) to the one obtained by using the heat kernel method~\cite{AlonsoIzquierdo:2011dy,AlonsoIzquierdo:2012tw}. The result from the heat kernel method has the value of VPE as $E_{\rm vac}=-0.73433$. Finally, in Figure \ref{fig:phase_b}, we show the phase shift for this model for $n=3$, thus confirming the Levinson's theorem.

\section{Conclusion}
\label{conc}
We have investigated the vacuum polarization energies (VPE) of kinks in the sinh-deformed $\phi^{4}$ and $\varphi^{6}$ models obtained from the polynomial $\phi^{4}$ and $\varphi^{6}$ models by a deformation procedure. We make use of spectral methods for computing the VPE in terms of scattering data for the quantum fluctuations about the classical kink. The models we used are not multiplicatively renormalizable. However, using the no-tadpole renormalization condition at one-loop order produce finite results for our calculations.
\begin{table}
	\centerline{
		\begin{tabular}{c |c c c|c|c|c}
			&\multicolumn{3}{c|}{bound state energies} 
			& $E_{\rm b.s}$ & $E_{\rm scat.}$ & $E_{\rm vac.}$\cr
			\hline
			polynomial $\phi^{4}$    & 0.0 & 1.225 &       & -0.802 & 0.354 & -0.448(-0.471) \cr
			\hline
			hyperbolic $\phi^{4}$    & 0.0 & 1.433 & 1.911 & -1.328 & 0.597 & -0.732(-0.734) \cr
	\end{tabular}}
	\caption{\label{t4}The bound state energies and VPEs of the hyperbolic $\phi^{4}$ model cf. Eq. (\ref{eq:hyper4})  and polynomial $\phi^{4}$ model cf. Eq (\ref{eq:phi4fhalve}). We have confirmed the VPE result using the Jost function formalism of Eq.~(\ref{eq:Jost}) which is indicated in the parenthesis of the last entry.}
\end{table}

Our results show that the sinh-deformed $\phi^{4}$ and $\varphi^{6}$ models show similar behaviors to their polynomial counterparts. In the case of the $\phi^{4}$ model they both have two bound states; the zero-mode and a vibrational shape mode. The narrowness of the scattering potential of the sinh-deformed $\phi^{4}$ model causes its vibrational mode frequency to be larger than that of its polynomial counterpart. In effect, the VPE of the sinh-deformed model is quite large. In the case of the $\varphi^{6}$ model with the dimensionless parameter $a=0$, the kink solutions of both models possesses non-equivalent vacua at spatial infinity. This leads to translational variance with respect to the center of the kink. The VPE may assume any negative value which then destabilizes the kink. The broadness of the scattering potential of the polynomial model in this case, causes its binding energy to be smaller than that of the sinh-deformed model. 

In the case of the hyperbolic $\phi^{4}$ model, we find its scattering potential to be broader than its polynomial counterpart. We also observe two vibrational modes for the hyperbolic model as compared to only one for its polynomial counterpart. In effect, the VPE of the hyperbolic $\phi^{4}$ model is smaller. The reported value of the VPE in this case, is  $E_{\rm vac}=-0.734$ which is in agreement with the value reported using the heat kernel method. The later approach makes use of $\zeta-$function regularization which requires truncation approximation.

It will be interesting to study the VPEs as a function of the kink-antikink potentials of the sinh-deformed models as well as the hyperbolic $\phi^{4}$ model. To do this requires to first substitute the configuration of the kink-antikink solutions that describe its scattering interactions to the Lagrangian density to obtain the classical kink potential. The VPE is then computed using the obtained potential. In deriving this configuration care must be taken in such a way that the configuration is not a solution to the stationary equations and additions must be made to avoid an imaginary VPE.

\section*{Acknowledgment}
We are grateful to H. Weigel for stimulating discussions and helpful comments on the manuscripts.

\end{document}